\documentclass[aps,prx,twocolumn,floats,showpacs,superscriptaddress,nofootinbib]{revtex4-1}
\usepackage{graphicx,epsfig}
\usepackage{times,bbm}
\usepackage{graphics,dcolumn,bm,float}
\usepackage{amssymb,amsmath,rotate,color}
\usepackage[title,titletoc,toc]{appendix}
\usepackage{mathtools}
\usepackage{booktabs}
\usepackage{tcolorbox}

\usepackage[pagebackref=false,colorlinks,linkcolor=magenta,citecolor=cyan,urlcolor=magenta]{hyperref}

\begin{document}
\unitlength 1 cm
\newcommand{\be}{\begin{equation}}
\newcommand{\ee}{\end{equation}}
\newcommand{\nn}{\nonumber}
\newcommand{\vk}{\vec k}
\newcommand{\vp}{\vec p}
\newcommand{\vq}{\vec q}
\newcommand{\vkp}{\vec {k'}}
\newcommand{\vpp}{\vec {p'}}
\newcommand{\vqp}{\vec {q'}}
\newcommand{\bsq}{{\boldsymbol{q}}}
\newcommand{\bsk}{{\boldsymbol{k}}}
\newcommand{\up}{\uparrow}
\newcommand{\down}{\downarrow}
\newcommand{\cdag}{c^{\dagger}}
\newcommand{\hlt}[1]{\textcolor{red}{#1}}
\newcommand{\ba}{\begin{align}}
\newcommand{\ea}{\end{align}}
\newcommand{\la}{\langle}
\newcommand{\ra}{\rangle}
\newcommand{\bearr}{\begin{eqnarray}}
\newcommand{\eearr}{\end{eqnarray}}
\newcommand{\eps}{\varepsilon}
\newcommand{\sgn}{\text{sgn}}
\newcommand{\bq}{{\boldsymbol{q}}}
\newcommand{\bk}{{\boldsymbol{k}}}
\newcommand{\bet}{{\boldsymbol{\eta}}}
\newcommand{\btau}{{\boldsymbol{\tau}}}
\newcommand{\bE}{{\boldsymbol{E}}}
\newcommand{\bB}{{\boldsymbol{B}}}
\newcommand{\bp}{{\boldsymbol{p}}}
\newcommand{\bv}{{\boldsymbol{v}}}
\newcommand{\br}{{\boldsymbol{r}}}
\newcommand{\pr}{\partial}
\newcommand{\bs}{\boldsymbol}
\newcommand{\bmt}{\left[\begin{matrix}}
\newcommand{\emt}{\end{matrix}\right]}

 \title{The $8Pmmn$ borophene sheet: A solid-state platform for space-time engineering}
   \author{T. Farajollahpour}
   \affiliation{Department of Physics, Sharif University of Technology, Tehran 11155-9161, Iran}
   
    \author{Z. Faraei}
   \affiliation{Department of Physics, Sharif University of Technology, Tehran 11155-9161, Iran}

 \author{S. A. Jafari}
 \email{akbar.jafari@gmail.com}
 \affiliation{Department of Physics, Sharif University of Technology, Tehran 11155-9161, Iran}
 \affiliation{Center of excellence for Complex Systems and Condensed Matter (CSCM), Sharif University of Technology, Tehran 1458889694, Iran}

\begin{abstract}
We construct the most generic Hamiltonian of the $8Pmmn$ structure of borophene sheet in presence of 
spin-orbit, as well as background electric and magnetic fields. 
In addition to spin and valley Hall effects, this structure offers a framework to conveniently manipulate the resulting "tilt" 
of the Dirac equation by applying appropriate electric fields. Therefore, the tilt can be made space-, as well as time-dependent. 
The border separating the low-field region with under-tilted Dirac fermions from the high-field region
with over-tilted Dirac fermions will correspond to a black-hole horizon. In this way, space-time dependent electric fields can be used to 
{\it design} the metric of the resulting space-time felt by electrons and holes satisfying the tilted Dirac equation. 
Our platform offers a way to generate analogues of gravitational waves by electric fields (instead of mass sources)
which can be detected in solid state spectroscopies as waves of enhanced superconducting correlations. 
\end{abstract}
\pacs{
}

\maketitle

\section{Introduction}
The dynamics of elementary particles is severely restricted by imposing the symmetry of vacuum, namely
the Lorentz symmetry on them. However elementary excitations in solid-state systems are mounted on a lattice.
As such, the low-energy (long wave-length) effective electronic degrees of freedom in solid-state systems are not obliged to satisfy 
the Lorentz symmetry, although they might do so, as in graphene sheets~\cite{Novoselov,katsnelson}
and 3+1 dimensional Dirac materials~\cite{Armitage,Wehling}. There are 230 possible symmetry structures on lattices~\cite{Dress}, 
some of which have non-symmorphic symmetry elements, namely elements that are a combination of point group operations
with fractional translation. The Bloch phase resulting from the shift can, for example, give rise to a class of
fermions dubbed nexus fermions which have no counterpart in the realm of elementary particles, as they boldly 
contradict the famous spin-statistics theorem according to which all fermions must have half-integer spins~\cite{Schwartz,Peskin}. 

\begin{figure}
	\centering
	\includegraphics[width =0.99 \linewidth]{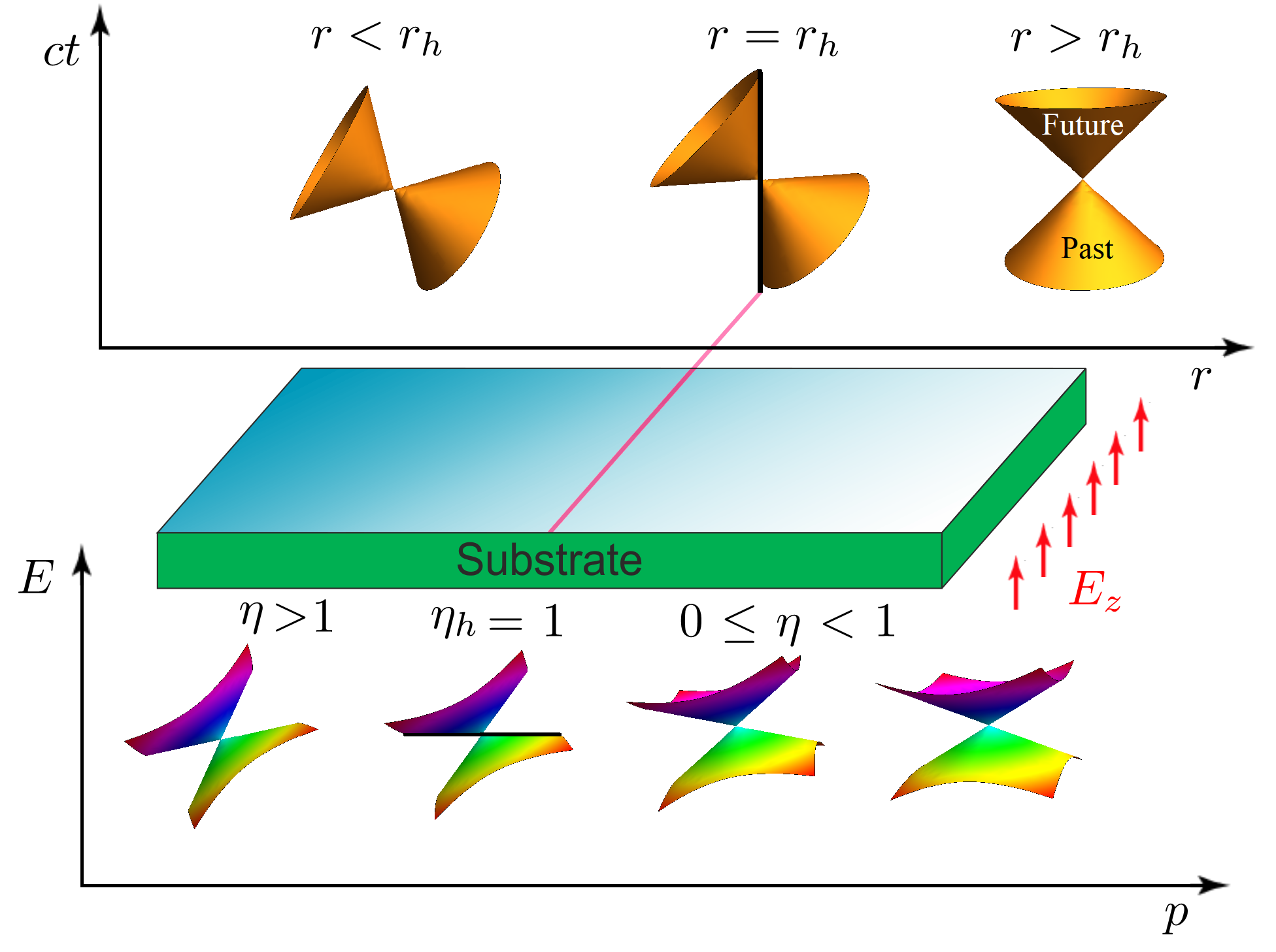}
	\caption{Schematic illustration of the low energy physics (solid state) analog of the black-hole horizon in $8Pmmn$ lattice. 
		The strong-field region will correspond to the interior of the black-hole.}
	\label{horizon}
\end{figure}

The non-symmorphic symmetry elements have at least one more interesting effect which is the subject of present work:
The resulting Dirac theory on the background of such lattices is specified by two velocity scales, (i) the major
velocity $v_F$ (that replaces the light velocity of high-energy relativistic theories), and (ii) the tilt velocity $v_t$. 
The ratio between these two parameters $\eta=v_t/v_F$ determines the type of tilted Dirac cone. 
The situation with $0<\eta<1$ ($1<\eta$) in Fig.~\ref{horizon}  is under-tilted (over-tilted).
Trying to tilt the Dirac equation in symmorphic structures such as graphene by applying strains will only produce very little tilt~\cite{Cabra,Mao}, 
while the pristine tilt in borophene is $\eta\sim 0.4$~\cite{Yao-Cat}. This signifies the importance
of underlying non-symmorphic lattice structure which serves to produce a substantial tilt even in the non-strained
structure. Not only that, as we will show in this paper, the peculiar symmetry of $8Pmmn$ borophene forces the background 
electric (and magnetic) fields to couple to electronic degrees of freedom in such a way that the tilt velocity $v_t$ can be directly tuned by the electric field. 

To set the stage for our work, let us start by the minimal form of tilted Dirac equation in two space dimensions~\cite{Tohyama2009,SaharTilt1,SaharTilt2},
\be
H=\hbar v_F \begin{pmatrix}  \eta k_x &  k_x-ik_y\\   k_x+ik_y &  \eta k_x  \end{pmatrix}=\hbar v_F (\eta k_x \tau_0+\bk.\bs{\tau}),
\label{nmatrixform}
\ee 
where $\tau_0$ is the unit $2\times 2$ matrix and $\tau_{i=1,2}$  are Pauli matrices. 
In order to make the physics transparent, we have used our freedom to choose a coordinate system such that the $k_x$ axis is along the tilt direction.
From the effective theory of $8Pmmn$-borophene it follows that the effective Hamiltonian around the other valley is obtained by 
$\eta\to -\eta$. So the valley degrees of freedom can be labeled by $\zeta=\pm1$ (see SM). 
The possible anisotropy of the Fermi velocity $v_F$~\cite{Suzumura}
can be removed by a rescaling of momenta (or coordinates) 
which will give rise to a constant Jacobian and does not alter the physics. 
The dispersion relation for this tilted Dirac cone Hamiltonian are given by,
\be
E_s(\bk)=  k (s+\eta \cos {\theta}_{\bk}) ~~~,~~~
\label{dispersion.eqn}
\ee
where $s=\pm 1$ refers to positive ($E_+$) and negative ($E_-$) energy states, and 
$\theta_{\bk} $ is polar angle of the two-dimensional wave vector, $\bk$, with respect to the $x$ axis. 

Following Volovik~\cite{Nissinen2017,Volovik2016Black,Volovik_2018}, let us view the dispersion of a tilted Dirac cone 
as a null-surface in a Painelev\'e-Gullstrand (PG) space-time,
\be
ds^2=-v_F^2 dt^2+(d\br-\bv_t dt)^2.
\ee
The $v_t$ can have arbitrary dependence on space-time coordinates. When $v_t$ is inversely proportional to the
radial coordinate, i.e. $\eta=v_t/v_F=r_h/r$ (where $r_h=2GM/v_F$ is the solid-state horizon radius),
the resulting PG metric will be a coordinate transformation of celebrated Schwartzchild space-time~\cite{Carroll} that remains
regular at the horizon~\cite{PG}. In our work $v_t$ can be arbitrary,
and we will show that it is solely controlled by the perpendicular electric field. Therefore in $8Pmmn$ borophene, the geometry of such a generic
PG space-time can be engineered via engineering the space-time profile of $v_t$ or equivalently $\eta$.   
The dispersion of massless particles in this space time is given by $g_{\mu\nu}k^\mu k^\nu=0$,
or equivalently, $(E-\bk.\bv_t)^2-v_F^2k^2=0$ which is equivalent to the dispersion relation~\eqref{dispersion.eqn}. 
When the tilt-velocity is allowed to depend on the radial coordinate as $v_t/v_F=r_h/r$, 
the over-tilt condition $v_t>v_F$ corresponds to $r<r_h$  which defines the balck-hole in this space-time geometry as in Fig.~\ref{horizon}.

Unlike existing proposals of black-hole physics \cite{Black-hole} in condensed matter systems based on liquid Helium~\cite{Artificial} 
or Bose-Einstein condensates in three space dimensions~\cite{macher,galitski}, 
which require very low-temperature or high pressures, the $8Pmmn$ borophene sheet offers a solid-state system at ambient conditions with additional tunability to explore black-hole physics on the table top. 
Recent solid-state proposals in 3+1 dimensions employ the external strain~\cite{njpGuan} or intrinsic inhomogeneity~\cite{HuangZnInS}
where a type-III Weyl material with $\eta=1$ (right at the horizon) can {\it slightly} change to either $\eta<1$ or $\eta>1$ (see Fig.~\ref{horizon}).
In our proposal, apart from dealing with a 2+1 dimensional space-time which makes our system a unique space-time laboratory on the
table-top, the electric field offers a wide tunability which is different from slight changes induced by strain. Furthermore, being
two-diemnsional allows to "pattern" the electric field to design arbitrary 2+1 dimensional space-time. 

\section{Results}
\subsection{Candidate material}
\begin{figure}
	\centering
	\includegraphics[width =0.99 \linewidth]{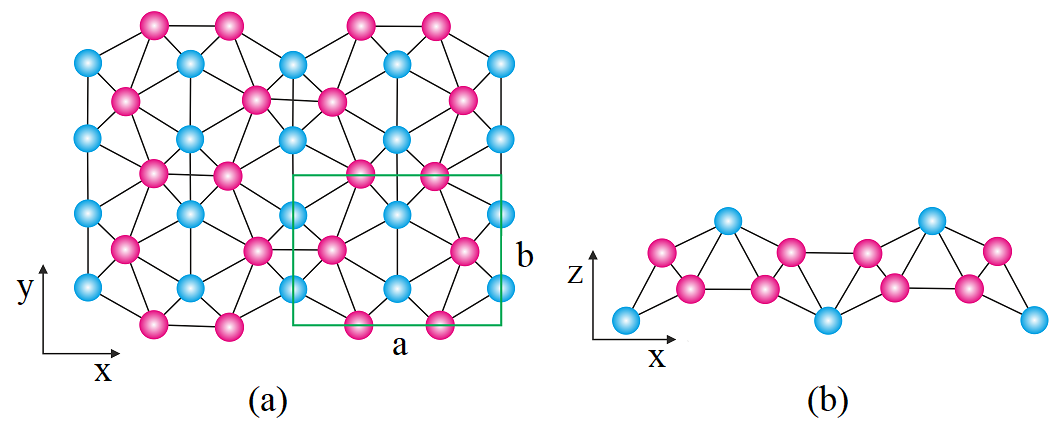}
	\caption{ Lattice structure of $8Pmmn$ borophene: 
		(a) top view (b) side view. Green rectangle is the unit cell. }
	\label{latticeRaw}
\end{figure}

As pointed out earlier, the non-symmorphic structure of the lattice is essential to produce substantial tilt
in the pristine form~\cite{MengFloquet}. Boron is the light element with the atomic number $Z=5$ which is in the left of Carbon in periodic table of elements. 
It has a number of structures~\cite{Ezawa} some of which are synthesized~\cite{exp1,exp2,exp3,exp4,exp5,exp6}. We will be dealing with 
the $8Pmmn$ structure shown in Fig.~\ref{latticeRaw} which is predicted to be stable~\cite{BeigiPRL,BeigiPRB,ZhoDFT,Littlewood}. 
Earlier layered material hosting tilted Dirac cone was the $\alpha$-BEDT organic conductor~\cite{Suzumura}. 
The common features of $8Pmmn$ borophene apart from hosting the tilted Dirac cones~\cite{Lozovik}, is that the low-energy degrees of freedom 
in both systems is built from molecular orbitals~\cite{ZhoDFT} rather than the atomic orbitals. However, their essential difference
is that the organic conductor is a layered material, while $8Pmmn$ borophene is an atom-thick two-dimensional material. 
In symmorphic structures such as graphene, the amount of tilt that can be extrinsically produced by applying an appropriate
strain on it is negligible. Furthermore, the electric field tunability of the tilt parameter of borophene is not available in graphene. 
Therefore we are left with only one suitable choice among the present three possible 2+1-dimensional tilted Dirac cone systems. 
Search for other 2+1-dimensional systems with substantial intrinsic tilt deformation of the Dirac equation
remains an open problem. 

\subsection{Effective Hamiltonian}
The most generic $8Pmmn$-invariant $4\times 4$ Hamiltonian in the basis of molecular orbitals,
$|\psi^c,\frac{1}{2}\rangle$, $|\psi^v,\frac{1}{2}\rangle$, $|\psi^c,-\frac{1}{2}\rangle$,$|\psi^v,-\frac{1}{2}\rangle$ 
is given by (see SM for details),
\begin{eqnarray}
&& H= f(\vk)\sigma_0\tau_0+ m(\vk)\sigma_0\tau_3 + t_0k_x\sigma_0\tau_2 \nonumber\\
&& +\Delta_{\rm KM} k_y\sigma_3\tau_1 \nn\\
&&+\sum_{i,j=1}^2 k_i\sigma_j\epsilon_{ij}  (\lambda^R_{0,i}\tau_0+\lambda^R_{3,i}\tau_3)\nn\\
&& + (\lambda_2\sigma_2 +\lambda_1 k_xk_y\sigma_1) \tau_2 \nn\\
&& +\sum_{i=1}^3 B_i\sigma_i g_{0,i}\tau_0 + M_i^{\rm int}g_{3,i} \tau_3 \label{effective.eqn}
\end{eqnarray}
where $f(\vk)=f_0+f_1k_x^2+f_2k_y^2$, $m(\vk)=m_1k_x^2+m_2k_y^2-m_0$ and we have used the explicit form 
of $\Gamma$ matrices in terms of direct product of Pauli matrices $\sigma$ (in spin-space) and $\tau$ (in molecular orbital space). 
$\Delta_{\rm KM}$ is spin-orbit coupling of the 
Kane-Mele~\cite{KaneMele}-type, $\lambda^R_{0,i}$ and $\lambda^R_{3,i}$ for $i=1,2$ are anisotropic
Rashba spin-orbit coupling. $\lambda^R_{0,i}$ is proportional to external electric field \cite{McDonald} while
$\lambda^R_{3,i}$ is similar to "buckling" term of Silicene structure~\cite{Ezawa}. $\lambda_1,\lambda_2$ are forms of spin-orbit coupling which are
specific to the $8Pmmn$ structure. The $\lambda$ couplings are proportional to the coordinate $z$ itself. 
Similarly, in the last line, we have two types of Zeeman coupling. The term containing $\tau_0$ are related
to coupling to external field $B$, and the term containing $\tau_3$ is due to internal exchange fields
which roots in the orbital angular momentum of the molecular orbitals involved. In both electric and magnetic field related terms, those couplings carrying 
the subscript $3$ which are coupled to $\tau_3$ arise from internal fields specific to $8Pmmn$ structure. The lack of symmetry under $z\to -z$ prevents them from vanishing.

\subsection{Spin Hall effect in borophene}
The first line of equation~\eqref{effective.eqn} is what gives the long-wave length limit in equation~\eqref{nmatrixform}. 
As long as only the first line is concerned, a pair of tilted Dirac cones at $\vk^\zeta=(0,\zeta k_{D})$  along the $\Gamma Y$ line
are obtained, 
where $k_D=\sqrt{m_0/m_2}$ and $\zeta=\pm$ is the valley index. The tilt is then controlled by $f_2$ parameter. 
The second line is related to spin-orbit coupling. Generically this term generates a gap in the tilted Dirac cone spectrum. 
The reason is simple, because in the first line $\tau_3$ and $\tau_2$ are already used, and the second line
being containing $\tau_1$ in two space dimensions will always generate a gap. 
To see this, let us define $\vp=\vk-\vk^\zeta$ and linearize the above Hamiltonian when only the first two lines are present:
\begin{eqnarray}
H= F_0\tau_0+F_1\sigma_3\tau_1 + F_2\tau_2 + F_3\tau_3 
\label{twolines.eqn}
\end{eqnarray}
where 
$F_0=2\zeta f_2k_{D} p_y$,
$F_1=\zeta \Delta_{\rm KM} k_D + \Delta_{\rm KM}p_y$,
$F_2= t_0p_x$ and $F_3= 2\zeta m_2 k_D p_y $. The spectrum of the above Hamiltonian for two eigenvalues $\sigma_3=\pm 1$ (for $\up$ and $\down$ states)
is given by 
\be
2\zeta f_2k_{D} p_y \pm \sqrt{\Delta_{\rm KM}^2(\zeta k_D + p_y)^2+(t_0p_x )^2+4(m_2 k_D p_y\sigma_3)^2}\nn
\ee
As can be seen, the tilt is controlled by $\zeta f_2 k_D$ and is therefore opposite for the 
two valleys. Moreover at Dirac nodes $\vk^\zeta$, we have $\vp=0$ and therefore the resulting
mass term is a {\it Haldane mass} and will be given by $\zeta \Delta_{\rm KM} k_D$ \cite{Haldane1988} 
which will give rise to SHE~\cite{Tokura2019,Jungwirth2012,Manchon2015}. Hydrogenation brings in some $sp^3$ component which
enhances the spin-orbit term ($\Delta_{\rm KM}$ in our case). The valley and spin dependence of the gap lies at the core of proposal by Kane and Mele~\cite{KaneMele}.
Observation of this effect for hydrogenated graphene has been discussed by Balakrishnan and coworkers~\cite{Balakrishnan2014}.
The further control parameter in the case of borophene is that in addition to the extrinsic enhancement of spin-orbit coupling ($\Delta_{\rm KM}$), 
one can also use the strain to manipulate $k_D$~\cite{Naumis2013,Manes2013}. 

A very important property of borophene in contrast to graphene is that, while in graphene
both topologically trivial Dirac mass and topologically non-trivial Haldane mass are allowed by symmetries of the lattice,
in the case of borophene, as long as it remains in the $8Pmmn$ symmetry group, the Haldane mass
is the {\it only} possible form of gap. To see this, let us try to add a gap opening
term proportional to $\tau_1$. If it is not of the $\sigma_3\tau_1$ form, then it must be proportional to $\sigma_0\tau_1=\Gamma_{45}$ which belongs to $B_{3u}$ representation (see SM) and is even with respect to
time-reversal.  But the basis functions in this representation are odd with respect to TR. 
This means that 
{\it only spin-orbit coupling is able to gap out the tilted Dirac cone spectrum of $8Pmmn$ borophene.}
Indeed this has been confirmed by {\it ab-initio} calculations. In the case of pristine borophene
where only intrinsic spin-orbit coupling exists, the spin-orbit induced gap is 
$\sim 0.03$ meV, while for hydrogenated borophene the buckling arising from $sp^3$ nature of the
bonds will give extrinsic contribution to spin-orbit coupling~\cite{NetoHC} which then generates
two orders of magnitude larger spin-orbit gap $\sim 2.25$ meV~\cite{Yao-Cat}. 
Note that, although the gap opening comes from $\sigma_3$ term, the two spin sub-bands corresponding 
to $\up$ and $\down$ spins remain degenerate as $\sigma_3^2=(\pm 1)^2=+1$. 
The effect of spin-orbit coupling in $8Pmmn$ group is different from e.g. $P4/nmm$ group where even spin-orbit coupling is not
able to gap out the resulting two-dimensional Dirac cone~\cite{YoungKane}. 

A further prediction of the first two lines of our effective Hamiltonian, equation~\eqref{effective.eqn}
which maybe relevant to $8Pmmn$ crystals other than borophene is that if a material in this group happen to have $m_0=0$, we will have $k_D=0$ 
and therefore the Dirac cone will move to $\Gamma$ pint. This will have two consequences: (i) 
The gap which is controlled by Haldane mass $\zeta \Delta_{\rm KM}k_D$ vanishes. (ii) The tilt which is controlled by
the $\zeta f_2 k_D$ combination also vanishes. Therefore as $m_0\to 0$, the two tilted gapped Dirac spectrum
with opposite tilt move towards each other, and collide at $\Gamma$ pint, whereby both gap and tilt are destroyed. 
Since the $\Delta_{\rm KM}$ term being related to spin-orbit coupling is non-zero for extrinsic or intrinsic reasons, 
the only way to diminish the Haldane mass will be to require $k_D=0$. But this will also diminish the tilt.
In this way, the SHE and tilt are locked to each other and always go hand in hand. This suggests a possible connection between
the tilt and the $Z_2$ index of the resulting SHE state.

\subsection{Electric field effects}
Now let us discuss the third and fourth line of our effective Hamiltonian, equation~\eqref{effective.eqn}.
The spin structure is of the Rashba form $\bk\times{\vec \sigma}.\vec z$ where $\vec z=z\hat z$ is a 
vector perpendicular to the crystal sheet. 
This term being proportional to the real space coordinate $z$ is related to a linear electrostatic profile which is equivalent to a constant electric field, $E_z$ perpendicular to the borophene sheet. Having a Rashba spin-orbit coupling form can be potentially useful in spintronic applications~\cite{Manchon2015}. 
The origin of the electric field can be extrinsic or intrinsic. 
The externally applied electric field couples to both conduction and valence states
on equal footing. Therefore it will be isotropic in $\tau$ space and hence will be coupled through
$\tau_0$. This accounts for the first term in the third line where Rashba parameters
$\lambda^R_{0,i}$ are introduced and they are related to external electric field by $E_{\rm ext}z\alpha_i=\lambda^R_{0,i}$
where $\alpha_i$ with $i=1,2$ accounts for the anisotropy of the crystal. In the isotropic approximation, this constant will not depend on the direction $i$. 
The second term $\lambda^R_{3,i}$ is coupled via $\tau_3$, meaning that it couples asymmetrically to molecular
orbital degrees of freedom forming the conduction and valence bands. This can be traced back to the lack of symmetry under $z\to -z$ of the crystal in Fig.~\ref{latticeRaw}  which gives rise to staggered polarization pattern.

This staggered polarization when projected in the space of low-energy molecular orbitals $|\psi^c\rangle$ and $|\psi^v\rangle$,
generates $\tau_3$ term as in the third line of equation~\eqref{effective.eqn}. Similar arguments apply to the fourth line.
Therefore couplings $\lambda^R_{3,i}$, $\lambda_2$ and $\lambda_1$ arise from internal polarization fields
and are fixed by materials parameters. Progress in the calculation of polarization for periodic solids~\cite{Resta2002,Vanderbilt}
can be employed to obtain {\it ab-initio} estimates of these couplings for $8Pmmn$ borophene. 

The matrix structure of the third line in the space of molecular orbitals is 
$\lambda^R_{0,i}\tau_0+\lambda^R_{3,i}\tau_3={\rm diag}(\lambda^R_{0,i}+\lambda^R_{3,i},\lambda^R_{0,i}-\lambda^R_{3,i})$. 
Since $\lambda^R_0$ can be externally tunned by applied electric field, it can be used to
tune either of the upper or lower diagonal components to zero. In this way the
the resulting (anisotropic) Rashba term will become orbital-selective~\cite{Xie2014}. 
A uniaxial strain is expected to distort the low-energy molecular orbitals whereby the
intrinsic $\lambda^R_{3,i}$ couplings can be changed~\cite{Manes2013,Naumis2013}.

\subsection{Tunable tilt: a tool for space-time engineering}
Now we are ready to discuss the most important message of our work which is the tunning of the tilt parameter by electric fields in  $8Pmmn$ borophene sheets. 
The role of all $\lambda$ terms in equation~\eqref{effective.eqn} is to generate spin-orbit gaps (see the next subsection). But
since Boron is a very light element, the intrinsic spin-orbit gaps are on the scale of $0.02$ meV~\cite{Yao-Cat}. 
Therefore we can ignore the mass terms. To look into the velocity scales in $x$-direction we set $p_y=0$ in equation~\eqref{linearized.eqn} to obtain,
\begin{eqnarray}
&&\varepsilon_{\tau}(p_x)=sp_x\sqrt{t_0^2+u^2+2\tau w^2 }\\
&&u^2=\lambda_1^2k_D^2+\left(\lambda_{3,1}^R\right)^2+\left(\lambda_{0,1}^R\right)^2\nn\\
&&w^4=\lambda_1^2k_D^2\left(\lambda_{3,1}^R\right)^2+t_0^2 \left[\left(\lambda_{0,1}^R \right)^2 + \left(\lambda_{3,1}^R\right)^2 \right]\nn
\end{eqnarray}
where $s=\pm 1$ are eigenvalues of $\sigma_3$ and $\tau=\pm 1$ refer to the eigenvalues of $\tau_3$ matrix. 
Now define the tilt and major velocity scale by $v_t=\frac{v_+ + v_-}{2},~~~v_F=\frac{v_+ - v_-}{2}$
from which we get,
in the limit of very large electric fields such that $\lambda^R_{0,1}\gg t_0$ we obtain,
\begin{eqnarray}
v_{tx} \approx (t_0^2+u^2)^{1/2},~~~~
v_{Fx} \approx \frac{w^2}{\sqrt{t_0^2 +u^2}}
\end{eqnarray}
In this limit the major Fermi velocity $v_{Fx}$ is saturated, while $v_{tx}$ remarkable is 
linearly controlled by the electric field. 
Similarly to investigate the velocity scales related to $y$ direction, we ignore the gap and set $p_x=0$ which gives,
\begin{eqnarray}
v_{ty}=2\zeta f_2 k_D+s \lambda_{0,2}^R,~~~~
v_{Fy}=2\zeta m_2 k_D + s \lambda_{3,2}^R
\end{eqnarray}
This again shows a linear dependence of the $v_{ty}$ to electric field, while the $v_{Fy}$ does not change 
by electric field. 

Therefore the major Fermi velocities $v_{Fi}$ determining the solid angle subtended by the Dirac cone are essentially
controlled by intrinsic parameters and intrinsic (albeit anisotropic) Rashba couplings $\lambda^R_{3,i}$, while
the corresponding tilt velocity $v_{ti}$ is controlled by external electric field $\lambda^R_{0,i}$. 
The intrinsic Rashba spin-orbit energy scales for pristine borophene are $\lambda^R_{3,i}\sim 10^{-2}$ meV. 
By hydrogenation and introducing $sp^3$ component, it can be enhanced up to $\sim 2$ meV~\cite{Yao-Cat}. At
energy scales well above these scales, the spin-orbit gap can be ignored, and we are essentially dealing with a gapless (two-space dimensional) Dirac node. 
When the external electric field is zero, the cone is given by the tilt parameter $\eta=3.4/8.0\sim 0.4$~\cite{Yao-Cat}. By increasing the external electric field, this value
will keep increasing. Beyond the point corresponding to $\eta_h=1$, it will be overtilted~\cite{Volovik2016,Nissinen2017}.
The reason we have used the subscript $h$ (for "horizon") rather than $c$ (for "critical") is to emphasize gravitational
analogy~\cite{Nissinen2017,Artificial,Volovik2016}. 

Therefore $8Pmmn$ borophene is a promising solid-state
platform where a background electric field can manipulate the tilt velocity scales. 
The ability to tune the tile of a Dirac cone is already interesting by itself. 
Larger tilt enhances the effect of Landau quantizations and
makes the ultra quantum limit much more accessible than the upright Dirac cone~\cite{Tohyama2009}. Also when
it comes to plasmon oscillations, the tilt gives rise to a kink in the plasmon spectrum~\cite{SaharTilt1,SaharTilt2}. 
An interesting amplification of magnetic fields by crossed electric fields can also be achieved in tilted Dirac cone
systems where the effective "tilt-boosted" magnetic fields felt in the two valleys are reciprocally related~\cite{Jafari2019}. 


Equipped with the gravitational analogy, $8Pmmn$ borophene can be used as a "black-hole on the table top"~\footnote{
	In the same spirit that graphene is thought of as CERN on the table top}: One can apply strong enough perpendicular electric field to a portion of borophene sample. The region with the strong field will correspond to overtilted Dirac cone, while the low-field region will be described by undertilted Dirac equation. The strong-field region will correspond to the interior of the black-hole, while the low-field region will correspond to the exterior of the black-hole as in Fig.~\ref{horizon}. 

Moving across the horizon in Fig.~\ref{horizon},
corresponds to the Lifshitz transition of the Dirac dispersion which will leave a signature in superconducting correlations~\cite{Li}. 
Letting the electric field profile oscillate with time will cause the horizon to oscillate. This will act like a 
source of gravitational waves which describes the oscillatory behavior of the metric.
Within Einstein equation, the "source" determining the metric is the mass content of space time,
while in our case the metric is determined by background electric field. 
Fixing a location for the tip of scanning tunneling microscope can detect that oscillations of the
metric in the form of oscillatory superconducting correlations. Any time a distortion of space-time in the form of a 
"tilt-hump" with $\eta\approx 1$ reaches the STM tip, it will detect an enhancement of superconducting correlations~\cite{Li}. 
Therefore the condensed matter analogue of "gravitational waves" of our simulated space-time are the waves of superconducting correlations. 

Another effect of the horizon in condensed matter applications would be that, the electron-hole pairs
created by e.g., sun light near the horizon will have some chance to enter the black hole. Even a small
in-plane electric field bias can encourage e.g. holes more than electrons to dive into black hole where
their future light cone is limits them to $\eta>1$ region. Therefore the horizon is a barrier for the recombination of electron-hole pairs. 
Given the two space dimensional nature of our system, this effect may find potential applications in solar cells where
reduction of electron-hole recombination is a merit.


\subsection{Valley Hall Effect}
Now let us discuss the physics of $\lambda$ terms which are Rashba-type and $8Pmmn$ generalizations of Rashba.
Typical scale of $\lambda^R_0$ terms arising from electric fields in Germanene is $\sim 10$ meV \cite{EzawaPhase} while in the 
intrinsic case $\Delta_{\rm KM}\sim 0.03$ meV~\cite{Yao-Cat}. 
Therefore let us ignore the Kane-Mele term, and focus on the $\lambda$-terms only. 
The essential role of $\lambda$ terms is to open spin-orbit gaps, whereby to generate Dirac mass for the fermions of equation~\eqref{nmatrixform}. 
The effective Hamiltonian around Dirac valley labeled by $\zeta=\pm 1$ is given by ($\Delta_{\rm KM}=0,B=0$)
\begin{eqnarray}
H=&& (F_0 +\lambda^R_{0,1} p_x\sigma_2-\lambda^R_{0,2} p_y\sigma_1-\zeta \lambda^R_{0,2} k_D \sigma_1)\tau_0 \nn\\
+&&(F_2+\zeta \lambda_1 k_D p_x\sigma_1+\lambda_2\sigma_2)\tau_2 \nn\\
+&&(F_3 + \lambda^R_{3,1}p_x\sigma_2 - \lambda^R_{3,2}p_y\sigma_1 - \zeta \lambda^R_{3,2} k_D \sigma_1 )\tau_3 
\label{linearized.eqn}
\end{eqnarray}
where $F_\mu$ coefficients are given below equation~\eqref{twolines.eqn}. 
Setting $p_x=p_y=0$, we obtain the "gap matrix" as
\be
g=-\zeta \lambda^R_{0,2} k_D \sigma_1\tau_0+\lambda_2\sigma_2\tau_2+\zeta \lambda^R_{3,2} k_D \sigma_1\tau_3. 
\label{gapmat.eqn}
\ee
Note that the only $\lambda^R_{0,2}$ and $\lambda^R_{3,2}$ appear in the gap matrix. 
The other two Rashba terms corresponding to $x$ direction, namely $\lambda^R_{0,1}$ and $\lambda^R_{3,1}$ appear
as coefficients of $p_x$ term. Their role is to renormalize the Fermi velocity and the velocity scale associated with the tilt which was discussed in previous next subsection. 
Corresponding to two eigenvalues $s=\pm 1$ of matrix $\sigma_3$, the eigenvalues of the above matrix are
\bearr
-d\pm \zeta \lambda^R_{3,2}k_D,~~~+d\pm \zeta \lambda^R_{3,2}k_D,~~~
d=\sqrt{\lambda_2^2+\left(\zeta\lambda^R_{0,2}k_D\right)^2}.
\eearr
\begin{figure}
	\centering
	\includegraphics[width =0.9900 \linewidth]{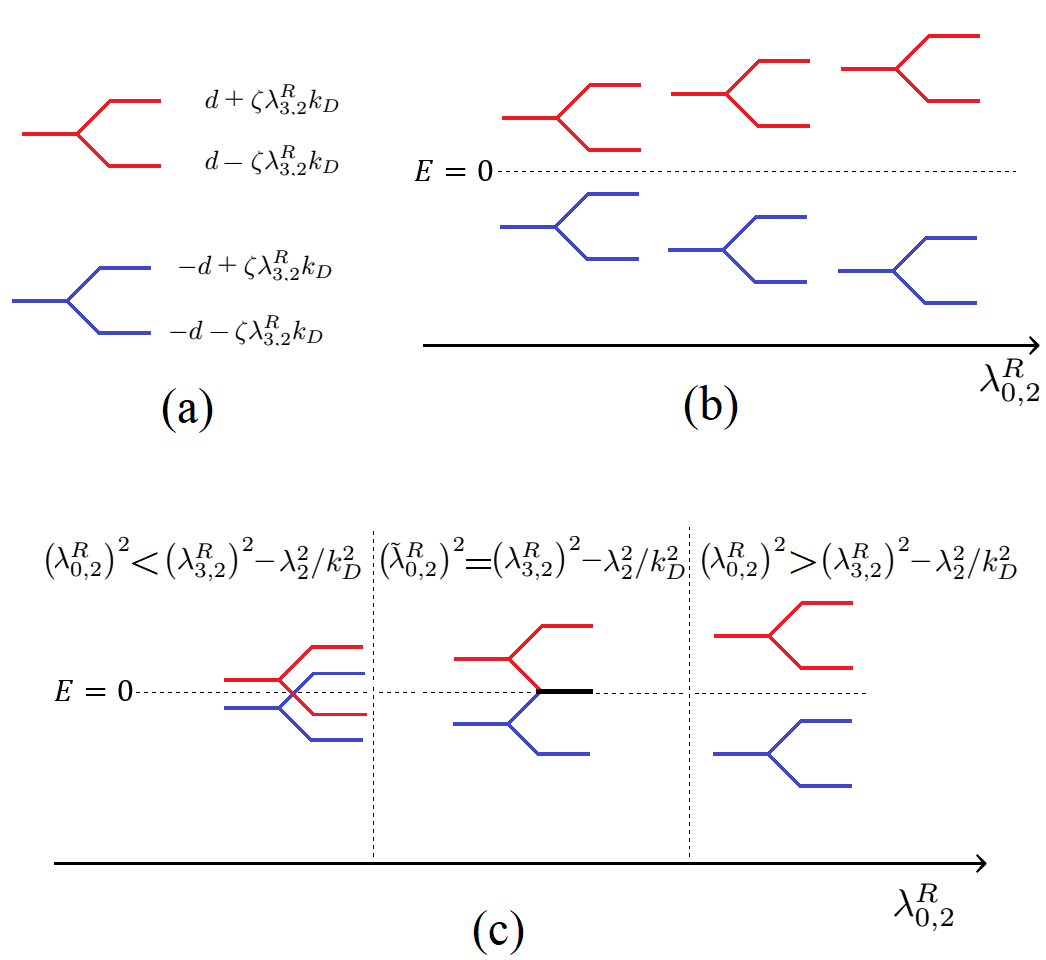}
	\caption{Schematic level diagram in the presence of external electric field. }
	\label{valleygap}
\end{figure}
Now there are two situations: (i) $|\lambda_2|>|\lambda^R_{3,2}|k_D$ and (ii) $|\lambda_2|<|\lambda^R_{3,2}|k_D$.
These are schematically shown in Fig.~\ref{valleygap} As can be seen from the schematic level diagram, the spectral gap in case (i)
is given by $2(d-\zeta\lambda^R_{3,2}k_D )$ (see Fig.~\ref{valleygap}.b). The valley symmetry is explicitly broken, which in turn will give rise
to valley Hall effect.  As can be seen by turning on the external field $\lambda^R_{0,2}$ and increasing
it, the gap always increases, and the valley asymmetry becomes less important. Therefore in case (i), the
valley Hall effect is decreased by turning on perpendicular electric field and increasing it.
Case (ii) is more interesting. In this case the gap will be given by $-2(d-\zeta\lambda^R_{3,2}k_D)$.
This quantity is positive when the external electric field is absent. At a critical value of 
$\left(\tilde\lambda^R_{0,2}\right)^2=\left(\lambda^R_{3,2}\right)^2-\lambda_2^2/k_D^2$
this gap vanishes, and beyond this point, the gap changes sign (see Fig.~\ref{valleygap}.c). In either case, the valley Hall effect
arising from the asymmetry between the Rashba gaps in the two valleys is present. But now by increasing
the externally applied perpendicular electric field, the sign of valley Hall effect can be changed.

\subsection{Magnetic field effects}
The last line contains the effect of Zeeman coupling. In table~II of supplementary material, we have
assumed that the "world" is composed of $8Pmmn$ crystal and the apparatus generating the
$\vec B$. In this way, under time-reversal $\vec B$ changes sign, and therefore to construct TR invariant Hamiltonian for the "world" (which is equivalent to breaking TR
of the "system") it has to couple to appropriate $\Gamma$ matrices which are odd with respect to TR.
In this way, we have obtained the last line of equation~\eqref{effective.eqn}. The first term
of the last line is similar to the standard Zeeman coupling which couples evenly to the molecular orbital degrees of freedom ($\propto \tau_0$). However, the second term
which containing $\tau_3$, asymmetrically  couples to orbital degrees of freedom. 
Similar to electric field case, the $g_{0,i}$ couplings are related to the externally applied field
(or exchange field) while $g_{3,i}$ couplings are intrinsic and arise from the orbital angular
momenta of the molecular orbitals involved. When the couplings are tunned to satisfy $g_{0,i}= \pm g_{3,i}$,
the Zeeman coupling will be orbital selective. 

To understand the physics of this line, let us assume that $\Delta_{\rm KM}\approx 0$ and $B\neq 0$, 
\begin{eqnarray}
H=(F_0 + \sum_{i=1}^3B_i\sigma_ig_{0,i})\tau_0 +F_2\tau_2 
+(F_3+ \sum_{i=1}^3M_i^{\rm int}\sigma_ig_{3,i})\tau_3\nn
\end{eqnarray}
The magnetic field has no effect on the Fermi and tilt velocities, as there is no term related to $B$ or $M$ in coefficients of 
$p_x$ and $p_y$ in Hamiltonian. Zeeman terms  generation gaps as,
\begin{eqnarray}
\varepsilon_g = s\sqrt{(B_i+\tau M^{\rm int}_i)^2}
\end{eqnarray} 
These gaps arising from  magnetic field interplay with the Kane-Mele gap and will enrich the phase diagram~\cite{EzawaPhase}. 


\section{Summary and Discussion}
We have obtained effective Hamiltonian of the non-symmorphic $8Pmmn$ borophene sheet which has a substantial intrinsic tilt in the 
spectrum of its Dirac fermions. Due to the non-symmorphic nature of the underlying lattice,
a perpendicular electric field couples to the system in such a way that it can tune the tilt parameter $v_t=\eta v_F$. 
The tilt parameter $\eta$ on the other hand can be regarded as a parameter in the effective space-time felt by 
the electrons and holes of the $8Pmmn$ graphene (Fig.~\ref{horizon}). In this analogy, the border between the high field region with overtilted Dirac cone
and a low-field region with undertilted Dirac cone will correspond to a horizon. For an electron moving near the horizon
the particle content of the states will be different from the one which is away from the horizon. Those approaching
the horizon will feel more particle fluctuations which correspond to enhanced superconducting correlations in a superconducting proximity set up. Letting the electric fields controlling the tilt $\eta$ to dance, will act like a 
source of "gravitational" waves which translate into waves of superconducting correlations. As such they can be detected 
as a wave of enhanced superconducting correlations in solid state spectroscopies. 

The ability to engineer the metric felt by electrons promotes our borophene system as a space-time simulator, 
albeit in 2+1 dimensions. Given that our real world is 3+1 dimensional, our proposal offers a unique solid-state
platform to explore 2+1 dimensional space-time which can not be found in cosmos. 

We also discussed valley Hall effect arising from the asymmetry $\eta\to -\eta$ of the two tilts. 
Extrinsically enhancing the spin-orbit coupling by e.g. hydrogenation will generate spin Hall effect
which would be perhaps comparable to the same effect in hydrogenated graphene~\cite{SOI3}.

\section{acknowledgements}
We wish to thank Mehdi Torabian, Shant Baghram, Mehdi Kargarian, Mojtaba Allaei and Abolhassan Vaezi for fruitful discussion.
T.F. appreciates Iman Ahmadabadi for useful discussion about the EBR. 
T.F. appreciates the financial support from Iran National Science Foundation (INSF) under post doctoral project no. 96015597. 
Z. F. was supported by Iran Science Elites Federation (ISEF) post doctoral fellowship. 
S. A. J. appreciates research deputy of Sharif University of Technology, grant no. G960214.

\appendix

\section{The nonsymmorphic $8Pmmn$ group}
\begin{figure}[b]
	\centering
	\includegraphics[width =0.85\linewidth]{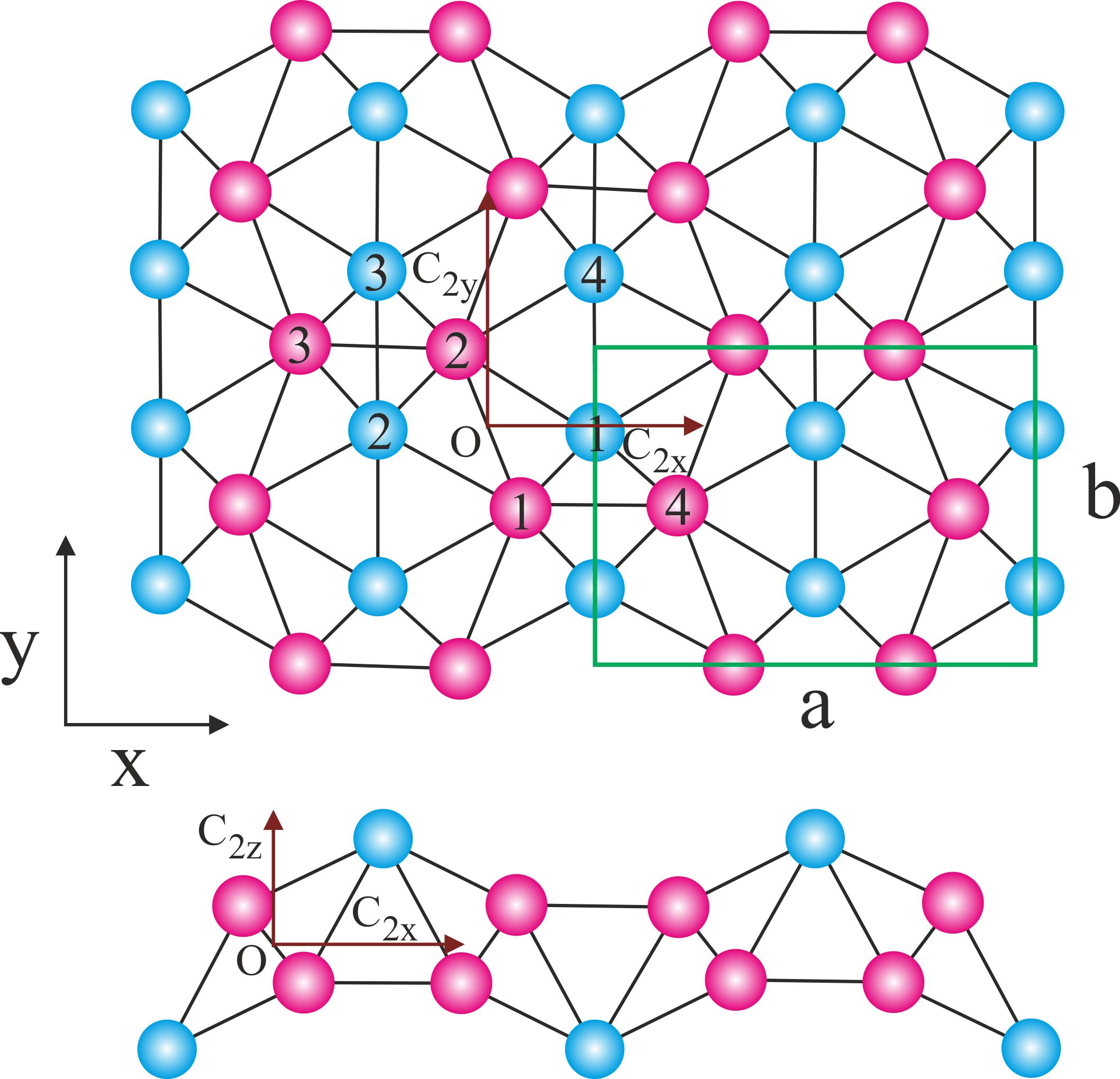}
	\caption{(Color online) Borophene lattice structure 
		(a) top view (b) side view. The screw symmetry axes for 
		$\tilde{c_{2x}}$, $\tilde{c_{2y}}$ and $\tilde{c_{2z}}$ 
		indicated by brown. Green rectangular is the selected unit cell.} 
	\label{lattice}
\end{figure}

To be self-contained, in this section we present details of the $8Pmmn$ group and its representations. 
The lattice structure of borophene is shown in Fig.~\ref{lattice}. 
The symmetry group of Borophene is $8Pmmn$, where the $8$ stands for the number of atoms
in the units cell. The $8Pmmn$ is point group number $59$~\cite{Bradley}. 
The generators (minimal set of elements from which all other members
of the group can be constructed) are given by~\cite{Bradley,Dress}, 
$\tilde{C}_{2x}=\{C_{2x}|\frac{a}{2}00 \}  ,\tilde{C}_{2y}= \{C_{2y}|0\frac{b}{2}0 \}  ,\tilde{I}=\{I|000\}$,
where the notation $\{C_{2x}|{\mathbf t}\}$ is a nonsymmorphic element meaning the two-fold rotation $C_{2x}$ around the $x$ axis 
is followed by a translation $\mathbf t$. All other symmetry operations of the $8Pmmn$ group can be constructed from the
above generators as follows:
$\tilde{C}_{2z}=\tilde{C}_{2x}\tilde{C}_{2y}=\{C_{2z}|\frac{a}{2}\frac{b}{2}0 \}$,
$\tilde{M}_z=\tilde{I}\tilde{C}_{2z}=\{M_z|\frac{a}{2}\frac{b}{2}0 \}$, 
$\tilde{M}_x=\tilde{I}\tilde{C}_{2x}=\{M_x|\frac{a}{2}00 \}$ and 
$\tilde{M}_y=\tilde{I}\tilde{C}_{2y}=\{M_y|0\frac{b}{2}0 \}$ \cite{Bradley}. 
This group has $8$ commuting elements and therefore being Abelian, admits $8$ one-dimensional
irreducible representations~\cite{BunkarChemist} given in Table.~\ref{Char-table}. 
Extensions beyond the double-lines are related to the double group which is obtained
by taking the spin of the electrons into account and that upon a $2\pi$ rotation,
the spinor goes into its negative~\cite{Dress}. The double group admits two-dimensional
representations which are not relevant to our minimal $4\times 4$ matrix representations
of the effective Hamiltonian of borophene. 

\begin{figure}
	\centering
	\includegraphics[width =0.5 \linewidth]{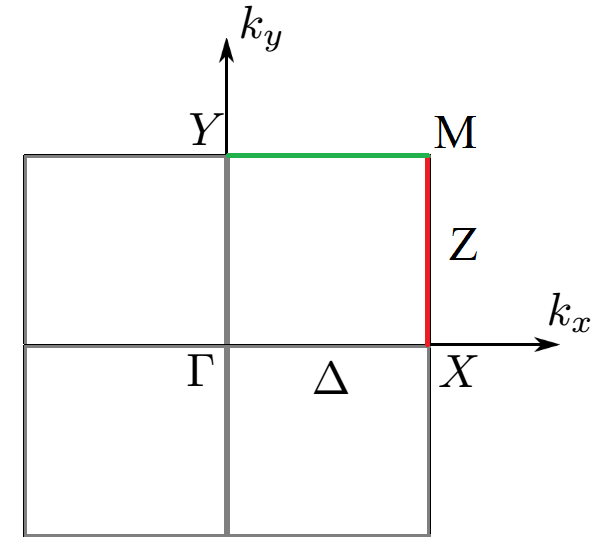}
	\caption{(Color online) Rectangular first Brillouin zone. }
	\label{BZ}
\end{figure}

Nonsymmorphic elements such as "gline planes and screw axes cause the
bands to {\em stick together} on the special surface lines and planes"~\cite{Kittel}. 
Following Kittel~\cite{Kittel} let us see how does it work for the $8Pmmn$ group. 
In the Brillouin zone of the system (Fig.~\ref{BZ}) let $\psi(x,y)$ be some basis wave function
on the boundary line $Z$ of the BZ. Then the screw rotation operation $\tilde C_{2x}$
acts as $\tilde{C}_{2x}\psi(x,y)=\psi(x+\frac{a}{2},-y)$ while the inversion $\tilde I$ is defined by 
the operation $\tilde I \psi(x,y)=\psi(-x,-y)$. On other hand $\tilde{M}_x$ acts as 
$\tilde{M}_x\psi(x,y) = \psi(-x+\frac{a}{2},y)$ which means,
\begin{eqnarray}\label{C1}
\tilde{C}_{2x}\psi(x,y) = \tilde{M}_x\tilde{I} \psi(x,y)
\end{eqnarray}
Now the argument by Kittel works by showing that assumption of non-degeneracy on $Z$ line
leads to contradiction~\cite{Kittel}. So let us assume that
on the $Z$ line, the representation is one dimensional.
From $\tilde{I}^2\psi(x,y)=\psi(x,y)$ it follows that $\tilde{I}\psi(x,y)=\pm \psi(x,y)$. 
The same story holds for 
$\tilde{M}_x\psi(x,y)=\pm \psi(x,y)$. Therefore by Eq.~\eqref{C1} we must have,
\begin{eqnarray}\label{C2}
&\tilde{C}_{2x}^2\psi(x,y)=\tilde{M}_x\tilde{I}\tilde{M}_x\tilde{I}\psi(x,y)\nonumber\\
&=(\pm1)^2(\pm1)^2\psi(x,y)=\psi(x,y)
\end{eqnarray} 
However from the very definition of $\tilde C_{2x}$ we have, 
\begin{eqnarray}
&\tilde{C}_{2x}^2\psi(x,y) = \tilde{C}_{2x} \psi(x+\frac{a}{2},-y) = \psi(x+a,+y) \nonumber\\
&=e^{ik_xa}\psi(x,y)=e^{i\pi}\psi(x,y) = -\psi(x,y)
\end{eqnarray}
\begin{table}
	\centering
	\label{Char-table}
	\caption{Character table of $8pmmn$ double group}
	\begin{tabular}{|l|l|l|l|l|l|l|l|l||l|l|}
		\hline
		& $E$ & $\tilde{c}_{2z}$ & $\tilde{c}_{2y}$  & $\tilde{c}_{2x}$ & $\tilde{I}$ & $\tilde{M}_{z}$ & $\tilde{M}_{y}$ & $\tilde{M}_{x}$ & $CE$ & $C\tilde{I}$  \\ \hline
		$A_g$    & 1 & 1 & 1 & 1 & 1 & 1 & 1 & 1 & 1 & 1 \\ \hline
		$A_u$    & 1 & 1 & 1 & 1 & -1 & -1 & -1 & -1 & 1 & -1 \\ \hline
		$B_{1g}$    & 1 & 1 & -1 & -1 & 1 & 1 & -1 & -1 & 1 & 1 \\ \hline
		$B_{1u}$    & 1 & 1 & -1 & -1 & -1 & -1 & 1 & 1 & 1 & -1 \\ \hline
		$B_{2g}$    & 1 & -1 & 1 & -1 & 1 & -1 & 1 & -1 & 1 & 1 \\ \hline
		$B_{2u}$    & 1 & -1 & 1 & -1 & -1 & 1 & -1 & 1 & 1 & -1 \\ \hline
		$B_{3g}$    & 1 & -1 & -1 & 1 & 1 & -1 & -1 & 1 & 1 & 1 \\ \hline
		$B_{3u}$    & 1 & -1 & -1 & 1 & -1 & 1 & 1 & -1 & 1 & -1 \\ \hline \hline
		$E_{g/2}$    & 2 & 0 & 0 & 0 & 2 & 0 & 0 & 0 & -2 & -2 \\ \hline
		$E_{u/2}$    & 2 & 0 & 0 & 0 & -2 & 0 & 0 & 0 & -2 & 2 \\ \hline
	\end{tabular}
\end{table}
which contradicts Eq.~\eqref{C2}. Note that in the the second line we have used 
the Bloch theorem and that on $Z$ line we have $k_xa=\pi$. 
Therefore, along the $XM=Z$ line of Fig.~\ref{BZ}, the irreducible representations can not
be one-dimensional. Now assuming that one of the states is $\psi$, again a similar argument 
by Kittel~\footnote{See page 214-215 of Kittel~\cite{Kittel}.}
shows that both $\psi$ and $\tilde C_{2x}T\psi$ where $T$ is the time-reversal operator, are degenerate. 
The same considerations apply to the $YM$ line. 
\begin{figure*}
	\centering
	\includegraphics[width =0.995 \linewidth]{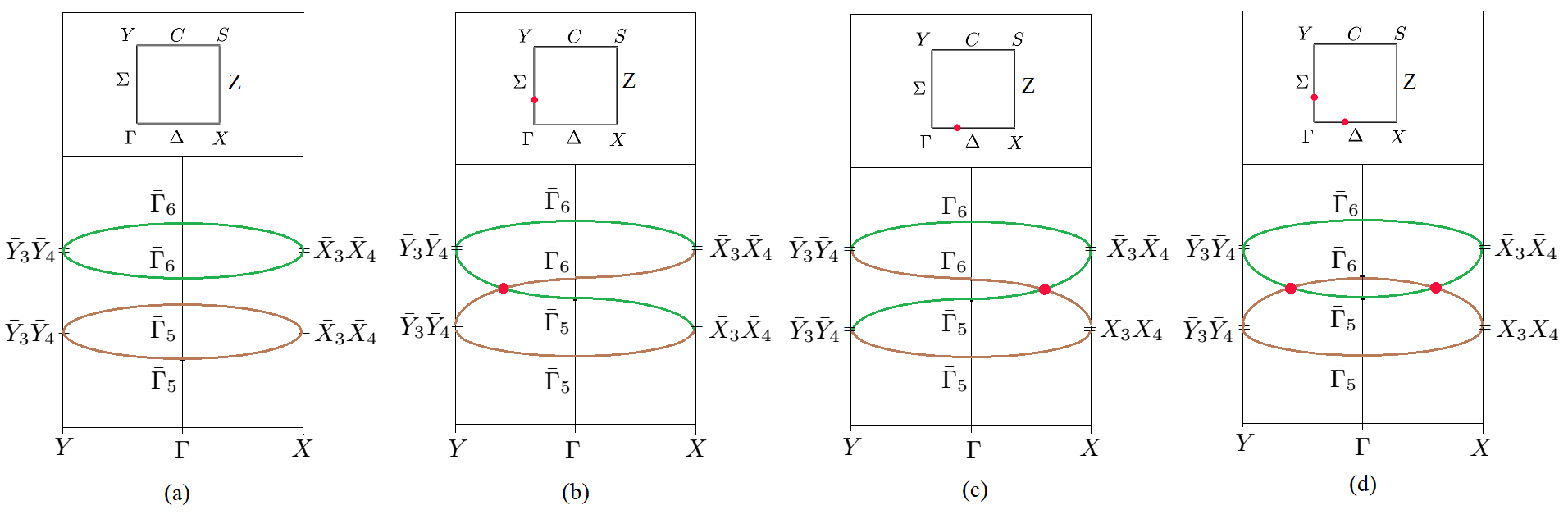}
	\caption{(Color online) Possible decomposition of elementary band representation 
		in the presence of time reversal symmetry labeled according to notations of 
		Bilbao crystallography server \cite{Bilbao}. (a) the disconnected components 
		of EBR correspond to the insulating phase of $8Pmmn$ group materials, (b) a Dirac 
		node points along the $\Gamma Y$  path, (c) a Dirac node
		along the $\Gamma X$  path and (d) a pair of Dirac points 
		along the $Y\Gamma X$ path in BZ.}
	\label{EBR}
\end{figure*}

\subsection{elementary band representation}
In this section by considering the Brillouin zone of the system, we obtain all the ways that are possible for energy 
bands to be connected in order to obtain realizable band structure \cite{EBRnature,EBRCano,EBRbradlyn}. 
In the band theory, the symmetry-enforced semimetal is realized when the number of     electrons is a fraction of the number of connections that forms an elementary band representation (EBR). The possible candidate 
for semimetallic materials are identified from EBRs~\cite{EBRnature,EBRCano,EBRbradlyn}. 

The possible decomposition of elementary band representation of 
the $8Pmmn$ group with time reversal symmetry is illustrated in 
Fig.~\ref{EBR}. As shown in Fig.~\ref{EBR}-a the disconnected components of EBR correspond to the insulating phase of $8Pmmn$ group materials. 
Fig.~\ref{EBR}-b and Fig.~\ref{EBR}-c show a graph with connected EBR which indicate the screw-symmetry and time reversal protected topological semimetal. 
In these two figures (Fig.~\ref{EBR}-b and Fig.~\ref{EBR}-c) Dirac nodes along the $\Gamma X$  and $\Gamma Y$ are obtained. 
According to EBR, there is yet another possibility shown in Fig.~\ref{EBR}-d
which corresponds to two pairs of Dirac points along the $Y \Gamma X$ path in BZ. 
However, as will become clear in the following sections and in agreement with {\it ab-initio}
calculations related to "borophene"~\cite{ZhoDFT}, the most general 
Hamiltonian constructed from irreducible representations will correspond Fig.~\ref{EBR}-b. 
There can be other possible materials with the same $8Pmmn$ which may realize
three other possibilities in Fig.~\ref{EBR}.

\section{Molecular orbitals and effective Hamiltonian}
The first known example of tilted Dirac cone material is the organic conductor $\alpha$-(BEDT-TTF)$_2$I$_3$,  
which is composed of molecular orbitals~\cite{Organic3}. It was noted by Zhou and coworkers that in the case of borophene
the charge density for the states at the bottom of the conduction band and top of valence band are enhanced on some bonds~\cite{ZhoDFT}.
On the other hand, {\it ab-initio} calculation of Ref.~\cite{Yao-Cat} shows that the eigenvalues of operators 
$(\tilde C_{2x},\tilde C_{2y},\tilde I)$ for bottom of conduction and top of valence band states are $(+1,+1,+1)$
and $(+1,-1,-1)$, respectively. 
This information is sufficient to let us construct most general molecular orbital consistent with the 
above eigenvalues. To do this, let us start from Fig.~\ref{lattice}. The pink atoms correspond to "inner" (${\rm I}$)
and blue atoms correspond to "ridge" (${\rm R}$) borons~\cite{Littlewood}. Both type of atoms are labeled by $1,2,3,4$. 
As can be seen in Fig.~\ref{lattice}, at the level of point group, we have following actions for the generators of $8Pmmn$ group:
$\tilde C_{2x}$ replace $(1^{\rm I}\leftrightarrow 3^{\rm I},2^{\rm I}\leftrightarrow 4^{\rm I})$, 
$(1^{\rm R}\leftrightarrow 2^{\rm R},3^{\rm R}\leftrightarrow 4^{\rm R})$.
For $\tilde C_{2y}$ we have 
$(1^{\rm I}\leftrightarrow 2^{\rm I},3^{\rm I}\leftrightarrow 4^{\rm I})$ and
$(1^{\rm R}\leftrightarrow 3^{\rm R},2^{\rm R}\leftrightarrow 4^{\rm R})$.
Finally $\tilde I$ acts as,
$(1^{\rm I}\leftrightarrow 2^{\rm I},3^{\rm I}\leftrightarrow 4^{\rm I})$ and
$(1^{\rm R}\leftrightarrow 2^{\rm R},3^{\rm R}\leftrightarrow 4^{\rm R})$. 
At the next level, depending on whether the relevant orbital in each of the above
positions is $p_x,p_y$ or $p_z$, we have:
\begin{eqnarray}
&\tilde{C}_{2x}\{p_x\} =p_x, \tilde{C}_{2x}\{p_{y,z}\} =-p_{y,z} \nonumber\\
&\tilde{C}_{2y}\{p_y\} =p_y, \tilde{C}_{2x}\{p_{x,z}\} =-p_{x,z} \nonumber\\
&\tilde{I}\{p_{x,y,z}\} =-p_{x,y,z}.\nn
\end{eqnarray}
It should be noted that the inversion center is located 
at the crossing point of two screw axis as plotted in Fig.~\ref{lattice}. 
Imposing the eigenvalues $(+1,+1,+1)$ and $(+1,-1,-1)$ for the conduction
and valence states, we obtain the most general molecular orbitals composing the bottom of conduction band ($|\Psi^c\rangle$) and
those at the top of valence band ($|\Psi^v\rangle$) as follows:
\begin{align}
&|\Psi^c_{\rm mol}\rangle\propto \label{cmol.eqn} \\
&\alpha \left(p_x^{\rm I1}-p_x^{\rm I2}+p_x^{\rm I3}-p_x^{\rm I4} \right)+\beta \left(p_y^{\rm R1}-p_y^{\rm R2}+p_y^{\rm R3}-p_y^{\rm R4} \right) \nonumber\\
&+\gamma \left(p_z^{\rm I1}-p_z^{\rm I2}-p_z^{\rm I3}+p_z^{\rm I4} \right)+\gamma' \left(p_z^{\rm R1}-p_z^{\rm R2}-p_z^{\rm R3}+p_z^{\rm R4} \right) \nonumber\\
&|\Psi^v_{\rm mol}\rangle\propto \label{vmol.eqn} \\
&\alpha' \left(p_x^{\rm I1}+p_x^{\rm I2}+p_x^{\rm I3}+p_x^{\rm I4} \right)+\alpha'' \left(p_x^{\rm R1}+p_x^{\rm R2}+p_x^{\rm R3}+p_x^{\rm R4} \right) \nonumber\\
&\gamma'' \left(p_z^{\rm I1}+p_z^{\rm I2}-p_z^{\rm I3}-p_z^{\rm I4} \right).\nn
\end{align}
As can be seen the eigenvalues of $(\tilde C_{2x},\tilde C_{2y},\tilde I)$ imply that the $p_y$ orbitals
are absent in the valence band states. This is in agreement with Ref.~\cite{Littlewood}. This reference further
suggests that the coefficient $\beta$ of the $p_y$ orbitals in the conduction band is also negligible. 
The coefficients can not be fixed by the symmetry argument. But this is already enough to let us 
construct the most general Hamiltonian compatible with the above eigenvalues. We have used the subscript
"mol" to emphasize the molecular orbital nature of degrees of freedom involved in the low-energy effective theory of borophene. 
This should be contrasted to a graphene sheet, where atomic $p_z$ orbitals form the low-energy electronic degrees of freedom. 

Eqns.~\eqref{cmol.eqn} and ~\eqref{vmol.eqn} allow us to construct the matrix representation 
of the generators (and hence all other members) of the $8Pmmn$ group. Let $\tau_i,i=1,2,3$ denote 
the Pauli matrices acting on the space of the above molecular orbitals. $\tau_0$ is the unit $2\times 2$ matrix.
Adding the spin structure, let $\sigma_i$ denote the Pauli matrices in the space of $\up$ and $\down$ states. Again
$\sigma_0$ will be the unit matrix in this space. In this basis we have the following representation,
\be
\tilde C_{2x}=i\sigma_1\tau_0,~~\tilde{C}_{2y}=i\sigma_2\tau_3,~~\tilde I=\sigma_0\tau_3,T=i\sigma_2\tau_0K
\label{opbasis.eqn}
\ee
where a tenser product $\otimes$ is understood. $T$ is the time-reversal, and $K$ is the complex conjugation. 
This representation is on the space of four states $|\Psi^c_{\rm mol}, \pm\frac{1}{2} \rangle $ and $|\Psi^v_{\rm mol}, \pm\frac{1}{2} \rangle $.
In this space, the most general $4\times4$ Hamiltonian can be written as, 
\begin{eqnarray}
H = d_0(k){\mathbbm 1} + \sum_i d_i(k)\Gamma_i + \sum_{ij}d_{ij}(k)\Gamma_{ij},
\end{eqnarray}
where $\mathbbm 1$ denotes the $4\times 4$ identity matrix and the $\Gamma$s are suitable basis in the space of
$4\times 4$ matrices. One possible explicit representation is given by~\cite{Liu2010},
\begin{eqnarray}
&\Gamma_1 = \sigma_x \otimes \tau_x \nonumber\\
&\Gamma_2 = \sigma_y \otimes \tau_x \nonumber\\
&\Gamma_3 = \sigma_z \otimes \tau_x \nonumber\\
&\Gamma_4 = 1 \otimes \tau_y \nonumber\\
&\Gamma_5 = 1 \otimes \tau_z \nonumber\\
&\Gamma_{ij} = [\sigma_i \otimes \tau_x, \sigma_j \otimes \tau_x ]/2i =\epsilon_{ijk}\sigma_k \otimes 1   \nonumber\\
&\Gamma_{i4}= [\sigma_i \otimes \tau_x, 1 \otimes \tau_y ]/2i =\sigma_i \otimes \tau_3  \nonumber\\
&\Gamma_{i5} = [\sigma_i \otimes \tau_x, 1 \otimes \tau_z ]/2i =-\sigma_i \otimes \tau_y\nonumber\\
&\Gamma_{45} = [1 \otimes \tau_y, 1 \otimes \tau_z ]/2i=1\otimes \tau_x 
\end{eqnarray}
Functions $d_0(k)$, $d_i(k)$ and $d_{ij}(k)$ are polynomials in $k$. Now using Eq.~\eqref{opbasis.eqn} 
one can construct the effect of all symmetry operators $g$ of the $8Pmmn$ group on the above set of $\Gamma$ matrices by
$g:\Gamma \to g\Gamma g^{-1}$.  

The properties of $\Gamma$ matrices under the generators of 
$8Pmmn$ group operators and time reversal symmetry operator 
$T$ are given in the following. For $\tilde C_{2x}$ we have,
\begin{eqnarray}
&\tilde{C}_{2x}\Gamma_1\tilde{C}_{2x}=\Gamma_1,~~\tilde{C}_{2x}\Gamma_2\tilde{C}_{2x}=-\Gamma_2,~~\tilde{C}_{2x}\Gamma_3\tilde{C}_{2x}=-\Gamma_3\nonumber \\
&\tilde{C}_{2x}\Gamma_4\tilde{C}_{2x}=\Gamma_4,~~\tilde{C}_{2x}\Gamma_5\tilde{C}_{2x}=\Gamma_5,~~\tilde{C}_{2x}\Gamma_{45}\tilde{C}_{2x}=\Gamma_{45} \nonumber\\
&\tilde{C}_{2x}\Gamma_{15}\tilde{C}_{2x}=\Gamma_{15},~~\tilde{C}_{2x}\Gamma_{25}\tilde{C}_{2x}=-\Gamma_{25},~~\tilde{C}_{2x}\Gamma_{35}\tilde{C}_{2x}=-\Gamma_{35} \nonumber\\
&\tilde{C}_{2x}\Gamma_{14}\tilde{C}_{2x}=\Gamma_{14},~~\tilde{C}_{2x}\Gamma_{24}\tilde{C}_{2x}=-\Gamma_{24},~~\tilde{C}_{2x}\Gamma_{34}\tilde{C}_{2x}=-\Gamma_{34} \nonumber\\
&\tilde{C}_{2x}\Gamma_{12}\tilde{C}_{2x}=-\Gamma_{12},~~\tilde{C}_{2x}\Gamma_{13}\tilde{C}_{2x}=-\Gamma_{13},~~\tilde{C}_{2x}\Gamma_{23}\tilde{C}_{2x}=\Gamma_{23} \nonumber\\
\end{eqnarray}
For $\tilde C_{2y}$ we obtain,
\begin{eqnarray}
&\tilde{C}_{2y}\Gamma_1\tilde{C}_{2y}=\Gamma_1,~~\tilde{C}_{2y}\Gamma_2\tilde{C}_{2y}=-\Gamma_2,~~\tilde{C}_{2y}\Gamma_3\tilde{C}_{2y}=\Gamma_3 \nonumber\\
&\tilde{C}_{2y}\Gamma_4\tilde{C}_{2y}=-\Gamma_4,~~\tilde{C}_{2y}\Gamma_5\tilde{C}_{2y}=\Gamma_5,~~\tilde{C}_{2y}\Gamma_{45}\tilde{C}_{2y}=-\Gamma_{45} \nonumber\\
&\tilde{C}_{2y}\Gamma_{15}\tilde{C}_{2y}=\Gamma_{15},~~\tilde{C}_{2y}\Gamma_{25}\tilde{C}_{2y}=-\Gamma_{25},~~\tilde{C}_{2y}\Gamma_{35}\tilde{C}_{2y}=\Gamma_{35} \nonumber\\
&\tilde{C}_{2y}\Gamma_{14}\tilde{C}_{2y}=-\Gamma_{14},~~\tilde{C}_{2y}\Gamma_{24}\tilde{C}_{2y}=\Gamma_{24},~~\tilde{C}_{2y}\Gamma_{34}\tilde{C}_{2y}=-\Gamma_{34} \nonumber\\
&\tilde{C}_{2y}\Gamma_{12}\tilde{C}_{2y}=-\Gamma_{12},~~\tilde{C}_{2y}\Gamma_{13}\tilde{C}_{2y}=\Gamma_{13},~~\tilde{C}_{2y}\Gamma_{23}\tilde{C}_{2y}=-\Gamma_{23} \nonumber\\
\end{eqnarray}
Under inversion they behave as,
\begin{eqnarray}
&\tilde{I}\Gamma_1\tilde{I}=-\Gamma_1,~~\tilde{I}\Gamma_2\tilde{I}=-\Gamma_2,~~\tilde{I}\Gamma_3\tilde{I}=-\Gamma_3 \\
&\tilde{I}\Gamma_4\tilde{I}=-\Gamma_4,~~\tilde{I}\Gamma_5\tilde{I}=\Gamma_5,~~\tilde{I}\Gamma_{45}\tilde{I}=-\Gamma_{45} \nonumber\\
&\tilde{I}\Gamma_{15}\tilde{I}=-\Gamma_{15},~~\tilde{I}\Gamma_{25}\tilde{I}=-\Gamma_{25},~~\tilde{I}\Gamma_{35}\tilde{I}=-\Gamma_{35} \nonumber\\
&\tilde{I}\Gamma_{14}\tilde{I}=\Gamma_{14},~~\tilde{I}\Gamma_{24}\tilde{I}=\Gamma_{24},~~\tilde{I}\Gamma_{34}\tilde{I}=\Gamma_{34} \nonumber\\
&\tilde{I}\Gamma_{12}\tilde{I}=\Gamma_{12},~~\tilde{I}\Gamma_{13}\tilde{I}=\Gamma_{13},~~\tilde{I}\Gamma_{23}\tilde{I}=\Gamma_{23} \nonumber
\end{eqnarray}
Finally under time-reversal they are transformed as,
\begin{eqnarray}
&T\Gamma_iT^{-1}=-\Gamma_i, ~~~i=1,2,3,4 \nonumber\\
&T\Gamma_5T^{-1}=\Gamma_5 \nonumber\\
&T\Gamma_{ij}T^{-1}=-\Gamma_{ij}~~~T\Gamma_{i4}T^{-1}=-\Gamma_{i4}\nonumber\\
&T\Gamma_{i5}T^{-1}=\Gamma_{i5}~~~i.j=1,2,3 \nonumber\\
&T\Gamma_{45}T^{-1}=\Gamma_{45}
\end{eqnarray}
Then using the character table~\ref{Char-table} the $\Gamma$ matrices can be classified 
in terms of the irreducible representations. We have summarized the result of this procedure in table~\ref{rep}. 
\begin{table}[b]
	\label{rep}
	\caption{Basis functions (polynomials up to third order) and $\Gamma$ 
		matrices transforming in every irreducible representation. The values of $T=\pm$ 
		denotes the signature under time reversal operation. As for the $B$ field itself, $T=-$
		indicates that the TR operates on the whole world.}
	\begin{tabular}{ccc}
		\hline
		\hline
		(basis;T)    &  representation  & \ \ ($\Gamma$ matrices;T) \\
		\hline\noalign{\smallskip}
		
		\{$1$,$k_x^2$,$k_y^2$ ;$+$\}        &   $A_g$    & \ \ \{I, $\Gamma_5$;$+$\} \\ 
		\{$k_xk_yz$;$+$ \}       &   $A_u$    & \ \ \{$\Gamma_{15}$;$+$\} \\
		\{$k_xk_y$;$+$\} &   $B_{1g}$     & \ \  ------  \\
		\{$B_z$;$-$ \}         &  $B_{1g}$    & \ \ \{$\Gamma_{12}$,$\Gamma_{34}$; -\} \\ 
		\{$z$  ;$+$\}        &   $B_{1u}$    & \ \ \{$\Gamma_{25}$;+\}  \\
		
		\{$B_y$,$zk_x$;$-$\}   &    $B_{2g}$    & \ \ \{$\Gamma_{13}$,$\Gamma_{24}$;$-$\}\\
		
		\{$k_y$,$k_x^2k_y$, $k_y^3$, $z^2k_y$;$-$\} &     $B_{2u}$   & \ \  \{$\Gamma_3$;$-$\}  \\
		
		\{$B_x$,$zk_y$;$-$\}  &     $B_{3g}$   & \ \ \{$\Gamma_{14}$, $\Gamma_{23}$;$-$ \} \\
		
		\{$k_x$, $k_xk_y^2$,$k^3_x$,$z^2k_x$;$-$\}  &      $B_{3u}$  & \ \ \{$\Gamma_4$;$-$\}   \\
		------ &      $A_u$  & \ \ \{$\Gamma_1$;$-$\}   \\
		------ &      $B_{1u}$  & \ \ \{$\Gamma_2$;$-$\}   \\
		------ &      $B_{2u}$  & \ \ \{$\Gamma_{35}$;$+$\}   \\
		------ &      $B_{3u}$  & \ \ \{$\Gamma_{45}$;$+$\}   \\
		
		\hline
	\end{tabular}
\end{table}

The middle column denotes the irreducible representations of the $8Pmmn$ group. It turns out that all the $16$ $\Gamma$-matrices
belong to one-dimensional representations of the $8Pmmn$ group. These are denoted in the right column, along with their signature
under TR operation. Corresponding basis functions up to third order polynomials along with their signature under TR are given in 
the left column. Since none of the matrices belongs to $B_{1g}$ representation, the corresponding entry is empty. In $A_u$ and $B_{1u}$
representations in rows number $10$ and $11$, the basis functions can only have "$+$" signature under the TR. So there is no basis function with "$-$" TR signature to
couple to $\Gamma_1$ and $\Gamma_2$ matrices. Similarly, in $B_{2u}$ and $B_{3u}$ irreducible representations of the last two rows, the basis functions 
are odd under TR, and there is no TR-even basis function to couple to $\Gamma_{35}$ and $\Gamma_{45}$ matrices. 
Now it is straightforward to construct invariant Hamiltonian: Simply multiply basis functions from the left column 
in their corresponding matrices in the right column. 

Therefore the most generic $8Pmmn$-invariant $4\times 4$ Hamiltonian 
$|\psi^c,\frac{1}{2}\rangle$, $|\psi^v,\frac{1}{2}\rangle$, $|\psi^c,-\frac{1}{2}\rangle$,$|\psi^v,-\frac{1}{2}\rangle$ basis 
is given by,
\begin{eqnarray}
&& H= f(\vk)\sigma_0\tau_0+ m(\vk)\sigma_0\tau_3 + t_0k_x\sigma_0\tau_2 \nonumber\\
&& +\Delta_{\rm KM} k_y\sigma_3\tau_1 \nn\\
&&+\sum_{i,j=1}^2 k_i\sigma_j\epsilon_{ij}  (\lambda^R_{0,i}\tau_0+\lambda^R_{3,i}\tau_3)\nn\\
&& + (\lambda_2\sigma_2 +\lambda_1 k_xk_y\sigma_1) \tau_2 \nn\\
&& +\sum_{i=1}^3 B_i\sigma_i g_{0,i}\tau_0 + M_i^{\rm int}g_{3,i} \tau_3 \label{effectiveSM.eqn}
\end{eqnarray}
where $f(\vk)=f_0+f_1k_x^2+f_2k_y^2$, $m(\vk)=m_1k_x^2+m_2k_y^2-m_0$ and we have used the explicit form 
of $\Gamma$ matrices in terms of direct product of $\sigma$ and $\tau$ matrices. $\Delta_{\rm KM}$ is spin-orbit coupling of the 
Kane-Mele~\cite{KaneMele}-type, $\lambda^R_{0,i}$ and $\lambda^R_{3,i}$ for $i=1,2$ are anisotropic
Rashba spin-orbit coupling. $\lambda^R_{0,i}$ is proportional to external electric field \cite{McDonald} while
$\lambda^R_{3,i}$ is similar to "buckling" term \cite{Ezawa}. $\lambda_1,\lambda_2$ are forms of spin-orbit coupling which are
specific to the $8Pmmn$ structure. All $\lambda$ couplings are proportional to the coordinate $z$ itself, signifying that
they are related to a linear potential profile. Those appearing along with $\tau_3$ are "staggered" fields, while those
evenly coupled to orbital degrees of freedom are "uniform" fields which can be extrinsically applied. 
Similarly, in the last line, we have two types of Zeeman coupling. The term proportional to $\tau_0$ are related
to coupling to external field $B_i$, and the term proportional to $\tau_3$ is due to internal exchange fields
which roots in the orbital angular momentum of the molecular orbitals involved. In both electric and magnetic field related terms, those couplings carrying index $3$ which are coupled to $\tau_3$ arise from internal fields specific to $8Pmmn$ structure. The lack of symmetry under $z\to -z$ prevents them from vanishing. 

\bibliographystyle{apsrev4-1}
\bibliography{Refs}

\begin{thebibliography}{66}%
\makeatletter
\providecommand \@ifxundefined [1]{%
 \@ifx{#1\undefined}
}%
\providecommand \@ifnum [1]{%
 \ifnum #1\expandafter \@firstoftwo
 \else \expandafter \@secondoftwo
 \fi
}%
\providecommand \@ifx [1]{%
 \ifx #1\expandafter \@firstoftwo
 \else \expandafter \@secondoftwo
 \fi
}%
\providecommand \natexlab [1]{#1}%
\providecommand \enquote  [1]{``#1''}%
\providecommand \bibnamefont  [1]{#1}%
\providecommand \bibfnamefont [1]{#1}%
\providecommand \citenamefont [1]{#1}%
\providecommand \href@noop [0]{\@secondoftwo}%
\providecommand \href [0]{\begingroup \@sanitize@url \@href}%
\providecommand \@href[1]{\@@startlink{#1}\@@href}%
\providecommand \@@href[1]{\endgroup#1\@@endlink}%
\providecommand \@sanitize@url [0]{\catcode `\\12\catcode `\$12\catcode
  `\&12\catcode `\#12\catcode `\^12\catcode `\_12\catcode `\%12\relax}%
\providecommand \@@startlink[1]{}%
\providecommand \@@endlink[0]{}%
\providecommand \url  [0]{\begingroup\@sanitize@url \@url }%
\providecommand \@url [1]{\endgroup\@href {#1}{\urlprefix }}%
\providecommand \urlprefix  [0]{URL }%
\providecommand \Eprint [0]{\href }%
\providecommand \doibase [0]{http://dx.doi.org/}%
\providecommand \selectlanguage [0]{\@gobble}%
\providecommand \bibinfo  [0]{\@secondoftwo}%
\providecommand \bibfield  [0]{\@secondoftwo}%
\providecommand \translation [1]{[#1]}%
\providecommand \BibitemOpen [0]{}%
\providecommand \bibitemStop [0]{}%
\providecommand \bibitemNoStop [0]{.\EOS\space}%
\providecommand \EOS [0]{\spacefactor3000\relax}%
\providecommand \BibitemShut  [1]{\csname bibitem#1\endcsname}%
\let\auto@bib@innerbib\@empty
\bibitem [{\citenamefont {Novoselov}\ \emph {et~al.}(2005)\citenamefont
  {Novoselov}, \citenamefont {Geim}, \citenamefont {Morozov}, \citenamefont
  {Jiang}, \citenamefont {Katsnelson}, \citenamefont {Grigorieva},
  \citenamefont {Dubonos}, \citenamefont {Firsov},\ and\ \citenamefont
  {AA}}]{Novoselov}%
  \BibitemOpen
  \bibfield  {author} {\bibinfo {author} {\bibfnamefont {K.~S.}\ \bibnamefont
  {Novoselov}}, \bibinfo {author} {\bibfnamefont {A.~K.}\ \bibnamefont {Geim}},
  \bibinfo {author} {\bibfnamefont {S.}~\bibnamefont {Morozov}}, \bibinfo
  {author} {\bibfnamefont {D.}~\bibnamefont {Jiang}}, \bibinfo {author}
  {\bibfnamefont {M.}~\bibnamefont {Katsnelson}}, \bibinfo {author}
  {\bibfnamefont {I.}~\bibnamefont {Grigorieva}}, \bibinfo {author}
  {\bibfnamefont {S.}~\bibnamefont {Dubonos}}, \bibinfo {author} {\bibnamefont
  {Firsov}}, \ and\ \bibinfo {author} {\bibnamefont {AA}},\ }\href@noop {}
  {\bibfield  {journal} {\bibinfo  {journal} {nature}\ }\textbf {\bibinfo
  {volume} {438}},\ \bibinfo {pages} {197} (\bibinfo {year}
  {2005})}\BibitemShut {NoStop}%
\bibitem [{\citenamefont {Katsnelson}\ and\ \citenamefont
  {Kat︠s︡nelʹson}(2012)}]{katsnelson}%
  \BibitemOpen
  \bibfield  {author} {\bibinfo {author} {\bibfnamefont {M.~I.}\ \bibnamefont
  {Katsnelson}}\ and\ \bibinfo {author} {\bibfnamefont {M.~I.}\ \bibnamefont
  {Kat︠s︡nelʹson}},\ }\href@noop {} {\emph {\bibinfo {title} {Graphene:
  carbon in two dimensions}}}\ (\bibinfo  {publisher} {Cambridge university
  press},\ \bibinfo {year} {2012})\BibitemShut {NoStop}%
\bibitem [{\citenamefont {Armitage}\ \emph {et~al.}(2018)\citenamefont
  {Armitage}, \citenamefont {Mele},\ and\ \citenamefont
  {Vishwanath}}]{Armitage}%
  \BibitemOpen
  \bibfield  {author} {\bibinfo {author} {\bibfnamefont {N.~P.}\ \bibnamefont
  {Armitage}}, \bibinfo {author} {\bibfnamefont {E.~J.}\ \bibnamefont {Mele}},
  \ and\ \bibinfo {author} {\bibfnamefont {A.}~\bibnamefont {Vishwanath}},\
  }\href {\doibase 10.1103/RevModPhys.90.015001} {\bibfield  {journal}
  {\bibinfo  {journal} {Rev. Mod. Phys.}\ }\textbf {\bibinfo {volume} {90}},\
  \bibinfo {pages} {015001} (\bibinfo {year} {2018})}\BibitemShut {NoStop}%
\bibitem [{\citenamefont {Wehling}\ \emph {et~al.}(2014)\citenamefont
  {Wehling}, \citenamefont {Black-Schaffer},\ and\ \citenamefont
  {Balatsky}}]{Wehling}%
  \BibitemOpen
  \bibfield  {author} {\bibinfo {author} {\bibfnamefont {T.}~\bibnamefont
  {Wehling}}, \bibinfo {author} {\bibfnamefont {A.~M.}\ \bibnamefont
  {Black-Schaffer}}, \ and\ \bibinfo {author} {\bibfnamefont {A.~V.}\
  \bibnamefont {Balatsky}},\ }\href@noop {} {\bibfield  {journal} {\bibinfo
  {journal} {Advances in Physics}\ }\textbf {\bibinfo {volume} {63}},\ \bibinfo
  {pages} {1} (\bibinfo {year} {2014})}\BibitemShut {NoStop}%
\bibitem [{\citenamefont {Dresselhaus}\ \emph {et~al.}(2007)\citenamefont
  {Dresselhaus}, \citenamefont {Dresselhaus},\ and\ \citenamefont
  {Jorio}}]{Dress}%
  \BibitemOpen
  \bibfield  {author} {\bibinfo {author} {\bibfnamefont {M.~S.}\ \bibnamefont
  {Dresselhaus}}, \bibinfo {author} {\bibfnamefont {G.}~\bibnamefont
  {Dresselhaus}}, \ and\ \bibinfo {author} {\bibfnamefont {A.}~\bibnamefont
  {Jorio}},\ }\href@noop {} {\emph {\bibinfo {title} {Group theory: application
  to the physics of condensed matter}}}\ (\bibinfo  {publisher} {Springer
  Science \& Business Media},\ \bibinfo {year} {2007})\BibitemShut {NoStop}%
\bibitem [{\citenamefont {Schwartz}(2014)}]{Schwartz}%
  \BibitemOpen
  \bibfield  {author} {\bibinfo {author} {\bibfnamefont {M.~D.}\ \bibnamefont
  {Schwartz}},\ }\href@noop {} {\emph {\bibinfo {title} {Quantum field theory
  and the standard model}}}\ (\bibinfo  {publisher} {Cambridge University
  Press},\ \bibinfo {year} {2014})\BibitemShut {NoStop}%
\bibitem [{\citenamefont {Peskin}(2018)}]{Peskin}%
  \BibitemOpen
  \bibfield  {author} {\bibinfo {author} {\bibfnamefont {M.~E.}\ \bibnamefont
  {Peskin}},\ }\href@noop {} {\emph {\bibinfo {title} {An introduction to
  quantum field theory}}}\ (\bibinfo  {publisher} {CRC Press},\ \bibinfo {year}
  {2018})\BibitemShut {NoStop}%
\bibitem [{\citenamefont {Cabra}\ \emph {et~al.}(2013)\citenamefont {Cabra},
  \citenamefont {Grandi}, \citenamefont {Silva},\ and\ \citenamefont
  {Sturla}}]{Cabra}%
  \BibitemOpen
  \bibfield  {author} {\bibinfo {author} {\bibfnamefont {D.~C.}\ \bibnamefont
  {Cabra}}, \bibinfo {author} {\bibfnamefont {N.~E.}\ \bibnamefont {Grandi}},
  \bibinfo {author} {\bibfnamefont {G.~A.}\ \bibnamefont {Silva}}, \ and\
  \bibinfo {author} {\bibfnamefont {M.~B.}\ \bibnamefont {Sturla}},\ }\href
  {\doibase 10.1103/PhysRevB.88.045126} {\bibfield  {journal} {\bibinfo
  {journal} {Phys. Rev. B}\ }\textbf {\bibinfo {volume} {88}},\ \bibinfo
  {pages} {045126} (\bibinfo {year} {2013})}\BibitemShut {NoStop}%
\bibitem [{\citenamefont {Mao}\ \emph {et~al.}(2011)\citenamefont {Mao},
  \citenamefont {Wang}, \citenamefont {Wei}, \citenamefont {Kaxiras},\ and\
  \citenamefont {Sodroski}}]{Mao}%
  \BibitemOpen
  \bibfield  {author} {\bibinfo {author} {\bibfnamefont {Y.}~\bibnamefont
  {Mao}}, \bibinfo {author} {\bibfnamefont {W.~L.}\ \bibnamefont {Wang}},
  \bibinfo {author} {\bibfnamefont {D.}~\bibnamefont {Wei}}, \bibinfo {author}
  {\bibfnamefont {E.}~\bibnamefont {Kaxiras}}, \ and\ \bibinfo {author}
  {\bibfnamefont {J.~G.}\ \bibnamefont {Sodroski}},\ }\href {\doibase
  10.1021/nn103153x} {\bibfield  {journal} {\bibinfo  {journal} {ACS Nano}\
  }\textbf {\bibinfo {volume} {5}},\ \bibinfo {pages} {1395} (\bibinfo {year}
  {2011})},\ \bibinfo {note} {pMID: 21222462},\ \Eprint
  {http://arxiv.org/abs/https://doi.org/10.1021/nn103153x}
  {https://doi.org/10.1021/nn103153x} \BibitemShut {NoStop}%
\bibitem [{\citenamefont {Fan}\ \emph {et~al.}(2018)\citenamefont {Fan},
  \citenamefont {Ma}, \citenamefont {Fu}, \citenamefont {Liu},\ and\
  \citenamefont {Yao}}]{Yao-Cat}%
  \BibitemOpen
  \bibfield  {author} {\bibinfo {author} {\bibfnamefont {X.}~\bibnamefont
  {Fan}}, \bibinfo {author} {\bibfnamefont {D.}~\bibnamefont {Ma}}, \bibinfo
  {author} {\bibfnamefont {B.}~\bibnamefont {Fu}}, \bibinfo {author}
  {\bibfnamefont {C.-C.}\ \bibnamefont {Liu}}, \ and\ \bibinfo {author}
  {\bibfnamefont {Y.}~\bibnamefont {Yao}},\ }\href {\doibase
  10.1103/PhysRevB.98.195437} {\bibfield  {journal} {\bibinfo  {journal} {Phys.
  Rev. B}\ }\textbf {\bibinfo {volume} {98}},\ \bibinfo {pages} {195437}
  (\bibinfo {year} {2018})}\BibitemShut {NoStop}%
\bibitem [{\citenamefont {Morinari}\ \emph {et~al.}(2009)\citenamefont
  {Morinari}, \citenamefont {Himura},\ and\ \citenamefont
  {Tohyama}}]{Tohyama2009}%
  \BibitemOpen
  \bibfield  {author} {\bibinfo {author} {\bibfnamefont {T.}~\bibnamefont
  {Morinari}}, \bibinfo {author} {\bibfnamefont {T.}~\bibnamefont {Himura}}, \
  and\ \bibinfo {author} {\bibfnamefont {T.}~\bibnamefont {Tohyama}},\ }\href
  {\doibase 10.1143/JPSJ.78.023704} {\bibfield  {journal} {\bibinfo  {journal}
  {J. Phys. Soc. Jpn.}\ }\textbf {\bibinfo {volume} {78}},\ \bibinfo {pages}
  {023704} (\bibinfo {year} {2009})}\BibitemShut {NoStop}%
\bibitem [{\citenamefont {Jalali-Mola}\ and\ \citenamefont
  {Jafari}(2018{\natexlab{a}})}]{SaharTilt1}%
  \BibitemOpen
  \bibfield  {author} {\bibinfo {author} {\bibfnamefont {Z.}~\bibnamefont
  {Jalali-Mola}}\ and\ \bibinfo {author} {\bibfnamefont {S.~A.}\ \bibnamefont
  {Jafari}},\ }\href {\doibase 10.1103/PhysRevB.98.195415} {\bibfield
  {journal} {\bibinfo  {journal} {Phys. Rev. B}\ }\textbf {\bibinfo {volume}
  {98}},\ \bibinfo {pages} {195415} (\bibinfo {year}
  {2018}{\natexlab{a}})}\BibitemShut {NoStop}%
\bibitem [{\citenamefont {Jalali-Mola}\ and\ \citenamefont
  {Jafari}(2018{\natexlab{b}})}]{SaharTilt2}%
  \BibitemOpen
  \bibfield  {author} {\bibinfo {author} {\bibfnamefont {Z.}~\bibnamefont
  {Jalali-Mola}}\ and\ \bibinfo {author} {\bibfnamefont {S.~A.}\ \bibnamefont
  {Jafari}},\ }\href {\doibase 10.1103/PhysRevB.98.235430} {\bibfield
  {journal} {\bibinfo  {journal} {Phys. Rev. B}\ }\textbf {\bibinfo {volume}
  {98}},\ \bibinfo {pages} {235430} (\bibinfo {year}
  {2018}{\natexlab{b}})}\BibitemShut {NoStop}%
\bibitem [{\citenamefont {Kajita}\ \emph
  {et~al.}(2014{\natexlab{a}})\citenamefont {Kajita}, \citenamefont {Nishio},
  \citenamefont {Tajima}, \citenamefont {Suzumura},\ and\ \citenamefont
  {Kobayashi}}]{Suzumura}%
  \BibitemOpen
  \bibfield  {author} {\bibinfo {author} {\bibfnamefont {K.}~\bibnamefont
  {Kajita}}, \bibinfo {author} {\bibfnamefont {Y.}~\bibnamefont {Nishio}},
  \bibinfo {author} {\bibfnamefont {N.}~\bibnamefont {Tajima}}, \bibinfo
  {author} {\bibfnamefont {Y.}~\bibnamefont {Suzumura}}, \ and\ \bibinfo
  {author} {\bibfnamefont {A.}~\bibnamefont {Kobayashi}},\ }\href@noop {}
  {\bibfield  {journal} {\bibinfo  {journal} {Journal of the Physical Society
  of Japan}\ }\textbf {\bibinfo {volume} {83}},\ \bibinfo {pages} {072002}
  (\bibinfo {year} {2014}{\natexlab{a}})}\BibitemShut {NoStop}%
\bibitem [{\citenamefont {Nissinen}\ and\ \citenamefont
  {Volovik}(2017)}]{Nissinen2017}%
  \BibitemOpen
  \bibfield  {author} {\bibinfo {author} {\bibfnamefont {J.}~\bibnamefont
  {Nissinen}}\ and\ \bibinfo {author} {\bibfnamefont {G.~E.}\ \bibnamefont
  {Volovik}},\ }\href {\doibase 10.1134/S0021364017070013} {\bibfield
  {journal} {\bibinfo  {journal} {JETP Letters}\ }\textbf {\bibinfo {volume}
  {105}},\ \bibinfo {pages} {447} (\bibinfo {year} {2017})}\BibitemShut
  {NoStop}%
\bibitem [{\citenamefont {Volovik}(2016{\natexlab{a}})}]{Volovik2016Black}%
  \BibitemOpen
  \bibfield  {author} {\bibinfo {author} {\bibfnamefont {G.~E.}\ \bibnamefont
  {Volovik}},\ }\href {\doibase 10.1134/S0021364016210050} {\bibfield
  {journal} {\bibinfo  {journal} {JETP Letters}\ }\textbf {\bibinfo {volume}
  {104}},\ \bibinfo {pages} {645} (\bibinfo {year}
  {2016}{\natexlab{a}})}\BibitemShut {NoStop}%
\bibitem [{\citenamefont {Volovik}(2018)}]{Volovik_2018}%
  \BibitemOpen
  \bibfield  {author} {\bibinfo {author} {\bibfnamefont {G.~E.}\ \bibnamefont
  {Volovik}},\ }\href {\doibase 10.3367/ufne.2017.01.038218} {\bibfield
  {journal} {\bibinfo  {journal} {Physics-Uspekhi}\ }\textbf {\bibinfo {volume}
  {61}},\ \bibinfo {pages} {89} (\bibinfo {year} {2018})}\BibitemShut {NoStop}%
\bibitem [{\citenamefont {Carroll}(2003)}]{Carroll}%
  \BibitemOpen
  \bibfield  {author} {\bibinfo {author} {\bibfnamefont {S.}~\bibnamefont
  {Carroll}},\ }\href@noop {} {\emph {\bibinfo {title} {Spacetime and Geometry:
  An Introduction to General Relativity}}}\ (\bibinfo  {publisher} {Pearson},\
  \bibinfo {year} {2003})\BibitemShut {NoStop}%
\bibitem [{\citenamefont {Martel}\ and\ \citenamefont {Poisson}(2001)}]{PG}%
  \BibitemOpen
  \bibfield  {author} {\bibinfo {author} {\bibfnamefont {K.}~\bibnamefont
  {Martel}}\ and\ \bibinfo {author} {\bibfnamefont {E.}~\bibnamefont
  {Poisson}},\ }\href {\doibase 10.1119/1.1336836} {\bibfield  {journal}
  {\bibinfo  {journal} {Am. J. Phys.}\ }\textbf {\bibinfo {volume} {69}},\
  \bibinfo {pages} {476} (\bibinfo {year} {2001})}\BibitemShut {NoStop}%
\bibitem [{\citenamefont {Curiel}(2019)}]{Black-hole}%
  \BibitemOpen
  \bibfield  {author} {\bibinfo {author} {\bibfnamefont {E.}~\bibnamefont
  {Curiel}},\ }\href {\doibase 10.1038/s41550-018-0602-1} {\bibfield  {journal}
  {\bibinfo  {journal} {Nature Astronomy}\ }\textbf {\bibinfo {volume} {3}},\
  \bibinfo {pages} {27} (\bibinfo {year} {2019})}\BibitemShut {NoStop}%
\bibitem [{\citenamefont {Novello}\ \emph {et~al.}(2002)\citenamefont
  {Novello}, \citenamefont {Visser},\ and\ \citenamefont
  {Volovik}}]{Artificial}%
  \BibitemOpen
  \bibfield  {author} {\bibinfo {author} {\bibfnamefont {M.}~\bibnamefont
  {Novello}}, \bibinfo {author} {\bibfnamefont {M.}~\bibnamefont {Visser}}, \
  and\ \bibinfo {author} {\bibfnamefont {G.~E.}\ \bibnamefont {Volovik}},\
  }\href@noop {} {\emph {\bibinfo {title} {Artificial black holes}}}\ (\bibinfo
   {publisher} {World Scientific},\ \bibinfo {year} {2002})\BibitemShut
  {NoStop}%
\bibitem [{\citenamefont {Macher}\ and\ \citenamefont
  {Parentani}(2009)}]{macher}%
  \BibitemOpen
  \bibfield  {author} {\bibinfo {author} {\bibfnamefont {J.}~\bibnamefont
  {Macher}}\ and\ \bibinfo {author} {\bibfnamefont {R.}~\bibnamefont
  {Parentani}},\ }\href {\doibase 10.1103/PhysRevA.80.043601} {\bibfield
  {journal} {\bibinfo  {journal} {Phys. Rev. A}\ }\textbf {\bibinfo {volume}
  {80}},\ \bibinfo {pages} {043601} (\bibinfo {year} {2009})}\BibitemShut
  {NoStop}%
\bibitem [{\citenamefont {Curtis}\ \emph {et~al.}(2018)\citenamefont {Curtis},
  \citenamefont {Refael},\ and\ \citenamefont {Galitski}}]{galitski}%
  \BibitemOpen
  \bibfield  {author} {\bibinfo {author} {\bibfnamefont {J.~B.}\ \bibnamefont
  {Curtis}}, \bibinfo {author} {\bibfnamefont {G.}~\bibnamefont {Refael}}, \
  and\ \bibinfo {author} {\bibfnamefont {V.}~\bibnamefont {Galitski}},\
  }\href@noop {} {\bibfield  {journal} {\bibinfo  {journal} {arXiv preprint
  arXiv:1801.01607}\ } (\bibinfo {year} {2018})}\BibitemShut {NoStop}%
\bibitem [{\citenamefont {Guan}\ \emph {et~al.}(2017)\citenamefont {Guan},
  \citenamefont {Yu}, \citenamefont {Liu}, \citenamefont {Liu}, \citenamefont
  {Dong}, \citenamefont {Lu}, \citenamefont {Yao},\ and\ \citenamefont
  {Yang}}]{njpGuan}%
  \BibitemOpen
  \bibfield  {author} {\bibinfo {author} {\bibfnamefont {S.}~\bibnamefont
  {Guan}}, \bibinfo {author} {\bibfnamefont {Z.-M.}\ \bibnamefont {Yu}},
  \bibinfo {author} {\bibfnamefont {Y.}~\bibnamefont {Liu}}, \bibinfo {author}
  {\bibfnamefont {G.-B.}\ \bibnamefont {Liu}}, \bibinfo {author} {\bibfnamefont
  {L.}~\bibnamefont {Dong}}, \bibinfo {author} {\bibfnamefont {Y.}~\bibnamefont
  {Lu}}, \bibinfo {author} {\bibfnamefont {Y.}~\bibnamefont {Yao}}, \ and\
  \bibinfo {author} {\bibfnamefont {S.~A.}\ \bibnamefont {Yang}},\ }\href
  {\doibase 10.1038/s41535-017-0026-7} {\bibfield  {journal} {\bibinfo
  {journal} {npj Quantum Materials}\ }\textbf {\bibinfo {volume} {2}},\
  \bibinfo {pages} {23} (\bibinfo {year} {2017})}\BibitemShut {NoStop}%
\bibitem [{\citenamefont {Huang}\ \emph {et~al.}(2018)\citenamefont {Huang},
  \citenamefont {Jin},\ and\ \citenamefont {Liu}}]{HuangZnInS}%
  \BibitemOpen
  \bibfield  {author} {\bibinfo {author} {\bibfnamefont {H.}~\bibnamefont
  {Huang}}, \bibinfo {author} {\bibfnamefont {K.-H.}\ \bibnamefont {Jin}}, \
  and\ \bibinfo {author} {\bibfnamefont {F.}~\bibnamefont {Liu}},\ }\href
  {\doibase 10.1103/PhysRevB.98.121110} {\bibfield  {journal} {\bibinfo
  {journal} {Phys. Rev. B}\ }\textbf {\bibinfo {volume} {98}},\ \bibinfo
  {pages} {121110} (\bibinfo {year} {2018})}\BibitemShut {NoStop}%
\bibitem [{\citenamefont {Liu}\ \emph {et~al.}(2019)\citenamefont {Liu},
  \citenamefont {Sun},\ and\ \citenamefont {Meng}}]{MengFloquet}%
  \BibitemOpen
  \bibfield  {author} {\bibinfo {author} {\bibfnamefont {H.}~\bibnamefont
  {Liu}}, \bibinfo {author} {\bibfnamefont {J.-T.}\ \bibnamefont {Sun}}, \ and\
  \bibinfo {author} {\bibfnamefont {S.}~\bibnamefont {Meng}},\ }\href@noop {}
  {\bibfield  {journal} {\bibinfo  {journal} {arXiv preprint arXiv:1901.10058}\
  } (\bibinfo {year} {2019})}\BibitemShut {NoStop}%
\bibitem [{\citenamefont {Ezawa}(2017)}]{Ezawa}%
  \BibitemOpen
  \bibfield  {author} {\bibinfo {author} {\bibfnamefont {M.}~\bibnamefont
  {Ezawa}},\ }\href {\doibase 10.1103/PhysRevB.96.035425} {\bibfield  {journal}
  {\bibinfo  {journal} {Phys. Rev. B}\ }\textbf {\bibinfo {volume} {96}},\
  \bibinfo {pages} {035425} (\bibinfo {year} {2017})}\BibitemShut {NoStop}%
\bibitem [{\citenamefont {Mannix}\ \emph {et~al.}(2015)\citenamefont {Mannix},
  \citenamefont {Zhou}, \citenamefont {Kiraly}, \citenamefont {Wood},
  \citenamefont {Alducin}, \citenamefont {Myers}, \citenamefont {Liu},
  \citenamefont {Fisher}, \citenamefont {Santiago}, \citenamefont {Guest},
  \citenamefont {Yacaman}, \citenamefont {Ponce}, \citenamefont {Oganov},
  \citenamefont {Hersam},\ and\ \citenamefont {Guisinger}}]{exp1}%
  \BibitemOpen
  \bibfield  {author} {\bibinfo {author} {\bibfnamefont {A.~J.}\ \bibnamefont
  {Mannix}}, \bibinfo {author} {\bibfnamefont {X.-F.}\ \bibnamefont {Zhou}},
  \bibinfo {author} {\bibfnamefont {B.}~\bibnamefont {Kiraly}}, \bibinfo
  {author} {\bibfnamefont {J.~D.}\ \bibnamefont {Wood}}, \bibinfo {author}
  {\bibfnamefont {D.}~\bibnamefont {Alducin}}, \bibinfo {author} {\bibfnamefont
  {B.~D.}\ \bibnamefont {Myers}}, \bibinfo {author} {\bibfnamefont
  {X.}~\bibnamefont {Liu}}, \bibinfo {author} {\bibfnamefont {B.~L.}\
  \bibnamefont {Fisher}}, \bibinfo {author} {\bibfnamefont {U.}~\bibnamefont
  {Santiago}}, \bibinfo {author} {\bibfnamefont {J.~R.}\ \bibnamefont {Guest}},
  \bibinfo {author} {\bibfnamefont {M.~J.}\ \bibnamefont {Yacaman}}, \bibinfo
  {author} {\bibfnamefont {A.}~\bibnamefont {Ponce}}, \bibinfo {author}
  {\bibfnamefont {A.~R.}\ \bibnamefont {Oganov}}, \bibinfo {author}
  {\bibfnamefont {M.~C.}\ \bibnamefont {Hersam}}, \ and\ \bibinfo {author}
  {\bibfnamefont {N.~P.}\ \bibnamefont {Guisinger}},\ }\href {\doibase
  10.1126/science.aad1080} {\bibfield  {journal} {\bibinfo  {journal}
  {Science}\ }\textbf {\bibinfo {volume} {350}},\ \bibinfo {pages} {1513}
  (\bibinfo {year} {2015})}\BibitemShut {NoStop}%
\bibitem [{\citenamefont {Zhong}\ \emph
  {et~al.}(2017{\natexlab{a}})\citenamefont {Zhong}, \citenamefont {Zhang},
  \citenamefont {Cheng}, \citenamefont {Feng}, \citenamefont {Li},
  \citenamefont {Sheng}, \citenamefont {Li}, \citenamefont {Meng},
  \citenamefont {Chen},\ and\ \citenamefont {Wu}}]{exp2}%
  \BibitemOpen
  \bibfield  {author} {\bibinfo {author} {\bibfnamefont {Q.}~\bibnamefont
  {Zhong}}, \bibinfo {author} {\bibfnamefont {J.}~\bibnamefont {Zhang}},
  \bibinfo {author} {\bibfnamefont {P.}~\bibnamefont {Cheng}}, \bibinfo
  {author} {\bibfnamefont {B.}~\bibnamefont {Feng}}, \bibinfo {author}
  {\bibfnamefont {W.}~\bibnamefont {Li}}, \bibinfo {author} {\bibfnamefont
  {S.}~\bibnamefont {Sheng}}, \bibinfo {author} {\bibfnamefont
  {H.}~\bibnamefont {Li}}, \bibinfo {author} {\bibfnamefont {S.}~\bibnamefont
  {Meng}}, \bibinfo {author} {\bibfnamefont {L.}~\bibnamefont {Chen}}, \ and\
  \bibinfo {author} {\bibfnamefont {K.}~\bibnamefont {Wu}},\ }\href
  {http://stacks.iop.org/0953-8984/29/i=9/a=095002} {\bibfield  {journal}
  {\bibinfo  {journal} {Journal of Physics: Condensed Matter}\ }\textbf
  {\bibinfo {volume} {29}},\ \bibinfo {pages} {095002} (\bibinfo {year}
  {2017}{\natexlab{a}})}\BibitemShut {NoStop}%
\bibitem [{\citenamefont {Zhong}\ \emph
  {et~al.}(2017{\natexlab{b}})\citenamefont {Zhong}, \citenamefont {Kong},
  \citenamefont {Gou}, \citenamefont {Li}, \citenamefont {Sheng}, \citenamefont
  {Yang}, \citenamefont {Cheng}, \citenamefont {Li}, \citenamefont {Wu},\ and\
  \citenamefont {Chen}}]{exp3}%
  \BibitemOpen
  \bibfield  {author} {\bibinfo {author} {\bibfnamefont {Q.}~\bibnamefont
  {Zhong}}, \bibinfo {author} {\bibfnamefont {L.}~\bibnamefont {Kong}},
  \bibinfo {author} {\bibfnamefont {J.}~\bibnamefont {Gou}}, \bibinfo {author}
  {\bibfnamefont {W.}~\bibnamefont {Li}}, \bibinfo {author} {\bibfnamefont
  {S.}~\bibnamefont {Sheng}}, \bibinfo {author} {\bibfnamefont
  {S.}~\bibnamefont {Yang}}, \bibinfo {author} {\bibfnamefont {P.}~\bibnamefont
  {Cheng}}, \bibinfo {author} {\bibfnamefont {H.}~\bibnamefont {Li}}, \bibinfo
  {author} {\bibfnamefont {K.}~\bibnamefont {Wu}}, \ and\ \bibinfo {author}
  {\bibfnamefont {L.}~\bibnamefont {Chen}},\ }\href {\doibase
  10.1103/PhysRevMaterials.1.021001} {\bibfield  {journal} {\bibinfo  {journal}
  {Phys. Rev. Materials}\ }\textbf {\bibinfo {volume} {1}},\ \bibinfo {pages}
  {021001} (\bibinfo {year} {2017}{\natexlab{b}})}\BibitemShut {NoStop}%
\bibitem [{\citenamefont {Feng}\ \emph {et~al.}(2016)\citenamefont {Feng},
  \citenamefont {Zhang}, \citenamefont {Zhong}, \citenamefont {Li},
  \citenamefont {Li}, \citenamefont {Li}, \citenamefont {Cheng}, \citenamefont
  {Meng}, \citenamefont {Chen},\ and\ \citenamefont {Wu}}]{exp4}%
  \BibitemOpen
  \bibfield  {author} {\bibinfo {author} {\bibfnamefont {B.}~\bibnamefont
  {Feng}}, \bibinfo {author} {\bibfnamefont {J.}~\bibnamefont {Zhang}},
  \bibinfo {author} {\bibfnamefont {Q.}~\bibnamefont {Zhong}}, \bibinfo
  {author} {\bibfnamefont {W.}~\bibnamefont {Li}}, \bibinfo {author}
  {\bibfnamefont {S.}~\bibnamefont {Li}}, \bibinfo {author} {\bibfnamefont
  {H.}~\bibnamefont {Li}}, \bibinfo {author} {\bibfnamefont {P.}~\bibnamefont
  {Cheng}}, \bibinfo {author} {\bibfnamefont {S.}~\bibnamefont {Meng}},
  \bibinfo {author} {\bibfnamefont {L.}~\bibnamefont {Chen}}, \ and\ \bibinfo
  {author} {\bibfnamefont {K.}~\bibnamefont {Wu}},\ }\href@noop {} {\bibfield
  {journal} {\bibinfo  {journal} {Nature chemistry}\ }\textbf {\bibinfo
  {volume} {8}},\ \bibinfo {pages} {563} (\bibinfo {year} {2016})}\BibitemShut
  {NoStop}%
\bibitem [{\citenamefont {Zhang}\ \emph
  {et~al.}(2016{\natexlab{a}})\citenamefont {Zhang}, \citenamefont {Penev},\
  and\ \citenamefont {Yakobson}}]{exp5}%
  \BibitemOpen
  \bibfield  {author} {\bibinfo {author} {\bibfnamefont {Z.}~\bibnamefont
  {Zhang}}, \bibinfo {author} {\bibfnamefont {E.~S.}\ \bibnamefont {Penev}}, \
  and\ \bibinfo {author} {\bibfnamefont {B.~I.}\ \bibnamefont {Yakobson}},\
  }\href@noop {} {\bibfield  {journal} {\bibinfo  {journal} {Nature chemistry}\
  }\textbf {\bibinfo {volume} {8}},\ \bibinfo {pages} {525} (\bibinfo {year}
  {2016}{\natexlab{a}})}\BibitemShut {NoStop}%
\bibitem [{\citenamefont {Zhang}\ \emph
  {et~al.}(2016{\natexlab{b}})\citenamefont {Zhang}, \citenamefont {Mannix},
  \citenamefont {Hu}, \citenamefont {Kiraly}, \citenamefont {Guisinger},
  \citenamefont {Hersam},\ and\ \citenamefont {Yakobson}}]{exp6}%
  \BibitemOpen
  \bibfield  {author} {\bibinfo {author} {\bibfnamefont {Z.}~\bibnamefont
  {Zhang}}, \bibinfo {author} {\bibfnamefont {A.~J.}\ \bibnamefont {Mannix}},
  \bibinfo {author} {\bibfnamefont {Z.}~\bibnamefont {Hu}}, \bibinfo {author}
  {\bibfnamefont {B.}~\bibnamefont {Kiraly}}, \bibinfo {author} {\bibfnamefont
  {N.~P.}\ \bibnamefont {Guisinger}}, \bibinfo {author} {\bibfnamefont {M.~C.}\
  \bibnamefont {Hersam}}, \ and\ \bibinfo {author} {\bibfnamefont {B.~I.}\
  \bibnamefont {Yakobson}},\ }\href@noop {} {\bibfield  {journal} {\bibinfo
  {journal} {Nano letters}\ }\textbf {\bibinfo {volume} {16}},\ \bibinfo
  {pages} {6622} (\bibinfo {year} {2016}{\natexlab{b}})}\BibitemShut {NoStop}%
\bibitem [{\citenamefont {Tang}\ and\ \citenamefont
  {Ismail-Beigi}(2007)}]{BeigiPRL}%
  \BibitemOpen
  \bibfield  {author} {\bibinfo {author} {\bibfnamefont {H.}~\bibnamefont
  {Tang}}\ and\ \bibinfo {author} {\bibfnamefont {S.}~\bibnamefont
  {Ismail-Beigi}},\ }\href {\doibase 10.1103/PhysRevLett.99.115501} {\bibfield
  {journal} {\bibinfo  {journal} {Phys. Rev. Lett.}\ }\textbf {\bibinfo
  {volume} {99}},\ \bibinfo {pages} {115501} (\bibinfo {year}
  {2007})}\BibitemShut {NoStop}%
\bibitem [{\citenamefont {Tang}\ and\ \citenamefont
  {Ismail-Beigi}(2010)}]{BeigiPRB}%
  \BibitemOpen
  \bibfield  {author} {\bibinfo {author} {\bibfnamefont {H.}~\bibnamefont
  {Tang}}\ and\ \bibinfo {author} {\bibfnamefont {S.}~\bibnamefont
  {Ismail-Beigi}},\ }\href {\doibase 10.1103/PhysRevB.82.115412} {\bibfield
  {journal} {\bibinfo  {journal} {Phys. Rev. B}\ }\textbf {\bibinfo {volume}
  {82}},\ \bibinfo {pages} {115412} (\bibinfo {year} {2010})}\BibitemShut
  {NoStop}%
\bibitem [{\citenamefont {Zhou}\ \emph {et~al.}(2014)\citenamefont {Zhou},
  \citenamefont {Dong}, \citenamefont {Oganov}, \citenamefont {Zhu},
  \citenamefont {Tian},\ and\ \citenamefont {Wang}}]{ZhoDFT}%
  \BibitemOpen
  \bibfield  {author} {\bibinfo {author} {\bibfnamefont {X.-F.}\ \bibnamefont
  {Zhou}}, \bibinfo {author} {\bibfnamefont {X.}~\bibnamefont {Dong}}, \bibinfo
  {author} {\bibfnamefont {A.~R.}\ \bibnamefont {Oganov}}, \bibinfo {author}
  {\bibfnamefont {Q.}~\bibnamefont {Zhu}}, \bibinfo {author} {\bibfnamefont
  {Y.}~\bibnamefont {Tian}}, \ and\ \bibinfo {author} {\bibfnamefont {H.-T.}\
  \bibnamefont {Wang}},\ }\href {\doibase 10.1103/PhysRevLett.112.085502}
  {\bibfield  {journal} {\bibinfo  {journal} {Phys. Rev. Lett.}\ }\textbf
  {\bibinfo {volume} {112}},\ \bibinfo {pages} {085502} (\bibinfo {year}
  {2014})}\BibitemShut {NoStop}%
\bibitem [{\citenamefont {Lopez-Bezanilla}\ and\ \citenamefont
  {Littlewood}(2016)}]{Littlewood}%
  \BibitemOpen
  \bibfield  {author} {\bibinfo {author} {\bibfnamefont {A.}~\bibnamefont
  {Lopez-Bezanilla}}\ and\ \bibinfo {author} {\bibfnamefont {P.~B.}\
  \bibnamefont {Littlewood}},\ }\href {\doibase 10.1103/PhysRevB.93.241405}
  {\bibfield  {journal} {\bibinfo  {journal} {Phys. Rev. B}\ }\textbf {\bibinfo
  {volume} {93}},\ \bibinfo {pages} {241405} (\bibinfo {year}
  {2016})}\BibitemShut {NoStop}%
\bibitem [{\citenamefont {Zabolotskiy}\ and\ \citenamefont
  {Lozovik}(2016)}]{Lozovik}%
  \BibitemOpen
  \bibfield  {author} {\bibinfo {author} {\bibfnamefont {A.~D.}\ \bibnamefont
  {Zabolotskiy}}\ and\ \bibinfo {author} {\bibfnamefont {Y.~E.}\ \bibnamefont
  {Lozovik}},\ }\href {\doibase 10.1103/PhysRevB.94.165403} {\bibfield
  {journal} {\bibinfo  {journal} {Phys. Rev. B}\ }\textbf {\bibinfo {volume}
  {94}},\ \bibinfo {pages} {165403} (\bibinfo {year} {2016})}\BibitemShut
  {NoStop}%
\bibitem [{\citenamefont {Kane}\ and\ \citenamefont {Mele}(2005)}]{KaneMele}%
  \BibitemOpen
  \bibfield  {author} {\bibinfo {author} {\bibfnamefont {C.~L.}\ \bibnamefont
  {Kane}}\ and\ \bibinfo {author} {\bibfnamefont {E.~J.}\ \bibnamefont
  {Mele}},\ }\href {\doibase 10.1103/PhysRevLett.95.226801} {\bibfield
  {journal} {\bibinfo  {journal} {Phys. Rev. Lett.}\ }\textbf {\bibinfo
  {volume} {95}},\ \bibinfo {pages} {226801} (\bibinfo {year}
  {2005})}\BibitemShut {NoStop}%
\bibitem [{\citenamefont {Min}\ \emph {et~al.}(2006)\citenamefont {Min},
  \citenamefont {Hill}, \citenamefont {Sinitsyn}, \citenamefont {Sahu},
  \citenamefont {Kleinman},\ and\ \citenamefont {MacDonald}}]{McDonald}%
  \BibitemOpen
  \bibfield  {author} {\bibinfo {author} {\bibfnamefont {H.}~\bibnamefont
  {Min}}, \bibinfo {author} {\bibfnamefont {J.~E.}\ \bibnamefont {Hill}},
  \bibinfo {author} {\bibfnamefont {N.~A.}\ \bibnamefont {Sinitsyn}}, \bibinfo
  {author} {\bibfnamefont {B.~R.}\ \bibnamefont {Sahu}}, \bibinfo {author}
  {\bibfnamefont {L.}~\bibnamefont {Kleinman}}, \ and\ \bibinfo {author}
  {\bibfnamefont {A.~H.}\ \bibnamefont {MacDonald}},\ }\href {\doibase
  10.1103/PhysRevB.74.165310} {\bibfield  {journal} {\bibinfo  {journal} {Phys.
  Rev. B}\ }\textbf {\bibinfo {volume} {74}},\ \bibinfo {pages} {165310}
  (\bibinfo {year} {2006})}\BibitemShut {NoStop}%
\bibitem [{\citenamefont {Haldane}(1988)}]{Haldane1988}%
  \BibitemOpen
  \bibfield  {author} {\bibinfo {author} {\bibfnamefont {F.~D.~M.}\
  \bibnamefont {Haldane}},\ }\href {\doibase 10.1103/PhysRevLett.61.2015}
  {\bibfield  {journal} {\bibinfo  {journal} {Phys. Rev. Lett.}\ }\textbf
  {\bibinfo {volume} {61}},\ \bibinfo {pages} {2015} (\bibinfo {year}
  {1988})}\BibitemShut {NoStop}%
\bibitem [{\citenamefont {Tokura}\ \emph {et~al.}(2019)\citenamefont {Tokura},
  \citenamefont {Yasuda},\ and\ \citenamefont {Tsukazaki}}]{Tokura2019}%
  \BibitemOpen
  \bibfield  {author} {\bibinfo {author} {\bibfnamefont {Y.}~\bibnamefont
  {Tokura}}, \bibinfo {author} {\bibfnamefont {K.}~\bibnamefont {Yasuda}}, \
  and\ \bibinfo {author} {\bibfnamefont {A.}~\bibnamefont {Tsukazaki}},\ }\href
  {\doibase 10.1038/s42254-018-0011-5} {\bibfield  {journal} {\bibinfo
  {journal} {Nature Reviews Physics}\ } (\bibinfo {year} {2019}),\
  10.1038/s42254-018-0011-5}\BibitemShut {NoStop}%
\bibitem [{\citenamefont {Jungwirth}\ \emph {et~al.}(2012)\citenamefont
  {Jungwirth}, \citenamefont {Wunderlich},\ and\ \citenamefont
  {Olejn{\'i}k}}]{Jungwirth2012}%
  \BibitemOpen
  \bibfield  {author} {\bibinfo {author} {\bibfnamefont {T.}~\bibnamefont
  {Jungwirth}}, \bibinfo {author} {\bibfnamefont {J.}~\bibnamefont
  {Wunderlich}}, \ and\ \bibinfo {author} {\bibfnamefont {K.}~\bibnamefont
  {Olejn{\'i}k}},\ }\href {https://doi.org/10.1038/nmat3279} {\bibfield
  {journal} {\bibinfo  {journal} {Nature Materials}\ }\textbf {\bibinfo
  {volume} {11}},\ \bibinfo {pages} {382 EP } (\bibinfo {year} {2012})},\
  \bibinfo {note} {review Article}\BibitemShut {NoStop}%
\bibitem [{\citenamefont {Manchon}\ \emph {et~al.}(2015)\citenamefont
  {Manchon}, \citenamefont {Koo}, \citenamefont {Nitta}, \citenamefont
  {Frolov},\ and\ \citenamefont {Duine}}]{Manchon2015}%
  \BibitemOpen
  \bibfield  {author} {\bibinfo {author} {\bibfnamefont {A.}~\bibnamefont
  {Manchon}}, \bibinfo {author} {\bibfnamefont {H.~C.}\ \bibnamefont {Koo}},
  \bibinfo {author} {\bibfnamefont {J.}~\bibnamefont {Nitta}}, \bibinfo
  {author} {\bibfnamefont {S.~M.}\ \bibnamefont {Frolov}}, \ and\ \bibinfo
  {author} {\bibfnamefont {R.~A.}\ \bibnamefont {Duine}},\ }\href
  {https://doi.org/10.1038/nmat4360} {\bibfield  {journal} {\bibinfo  {journal}
  {Nature Materials}\ }\textbf {\bibinfo {volume} {14}},\ \bibinfo {pages} {871
  EP } (\bibinfo {year} {2015})},\ \bibinfo {note} {review Article}\BibitemShut
  {NoStop}%
\bibitem [{\citenamefont {Balakrishnan}\ \emph {et~al.}(2014)\citenamefont
  {Balakrishnan}, \citenamefont {Koon}, \citenamefont {Avsar}, \citenamefont
  {Ho}, \citenamefont {Lee}, \citenamefont {Jaiswal}, \citenamefont {Baeck},
  \citenamefont {Ahn}, \citenamefont {Ferreira}, \citenamefont {Cazalilla},
  \citenamefont {Neto},\ and\ \citenamefont {{\"O}zyilmaz}}]{Balakrishnan2014}%
  \BibitemOpen
  \bibfield  {author} {\bibinfo {author} {\bibfnamefont {J.}~\bibnamefont
  {Balakrishnan}}, \bibinfo {author} {\bibfnamefont {G.~K.~W.}\ \bibnamefont
  {Koon}}, \bibinfo {author} {\bibfnamefont {A.}~\bibnamefont {Avsar}},
  \bibinfo {author} {\bibfnamefont {Y.}~\bibnamefont {Ho}}, \bibinfo {author}
  {\bibfnamefont {J.~H.}\ \bibnamefont {Lee}}, \bibinfo {author} {\bibfnamefont
  {M.}~\bibnamefont {Jaiswal}}, \bibinfo {author} {\bibfnamefont {S.-J.}\
  \bibnamefont {Baeck}}, \bibinfo {author} {\bibfnamefont {J.-H.}\ \bibnamefont
  {Ahn}}, \bibinfo {author} {\bibfnamefont {A.}~\bibnamefont {Ferreira}},
  \bibinfo {author} {\bibfnamefont {M.~A.}\ \bibnamefont {Cazalilla}}, \bibinfo
  {author} {\bibfnamefont {A.~H.~C.}\ \bibnamefont {Neto}}, \ and\ \bibinfo
  {author} {\bibfnamefont {B.}~\bibnamefont {{\"O}zyilmaz}},\ }\href
  {https://doi.org/10.1038/ncomms5748} {\bibfield  {journal} {\bibinfo
  {journal} {Nature Communications}\ }\textbf {\bibinfo {volume} {5}},\
  \bibinfo {pages} {4748 EP } (\bibinfo {year} {2014})},\ \bibinfo {note}
  {article}\BibitemShut {NoStop}%
\bibitem [{\citenamefont {Oliva-Leyva}\ and\ \citenamefont
  {Naumis}(2013)}]{Naumis2013}%
  \BibitemOpen
  \bibfield  {author} {\bibinfo {author} {\bibfnamefont {M.}~\bibnamefont
  {Oliva-Leyva}}\ and\ \bibinfo {author} {\bibfnamefont {G.~G.}\ \bibnamefont
  {Naumis}},\ }\href {\doibase 10.1103/PhysRevB.88.085430} {\bibfield
  {journal} {\bibinfo  {journal} {Phys. Rev. B}\ }\textbf {\bibinfo {volume}
  {88}},\ \bibinfo {pages} {085430} (\bibinfo {year} {2013})}\BibitemShut
  {NoStop}%
\bibitem [{\citenamefont {Ma\~nes}\ \emph {et~al.}(2013)\citenamefont
  {Ma\~nes}, \citenamefont {de~Juan}, \citenamefont {Sturla},\ and\
  \citenamefont {Vozmediano}}]{Manes2013}%
  \BibitemOpen
  \bibfield  {author} {\bibinfo {author} {\bibfnamefont {J.~L.}\ \bibnamefont
  {Ma\~nes}}, \bibinfo {author} {\bibfnamefont {F.}~\bibnamefont {de~Juan}},
  \bibinfo {author} {\bibfnamefont {M.}~\bibnamefont {Sturla}}, \ and\ \bibinfo
  {author} {\bibfnamefont {M.~A.~H.}\ \bibnamefont {Vozmediano}},\ }\href
  {\doibase 10.1103/PhysRevB.88.155405} {\bibfield  {journal} {\bibinfo
  {journal} {Phys. Rev. B}\ }\textbf {\bibinfo {volume} {88}},\ \bibinfo
  {pages} {155405} (\bibinfo {year} {2013})}\BibitemShut {NoStop}%
\bibitem [{\citenamefont {Castro~Neto}\ and\ \citenamefont
  {Guinea}(2009)}]{NetoHC}%
  \BibitemOpen
  \bibfield  {author} {\bibinfo {author} {\bibfnamefont {A.~H.}\ \bibnamefont
  {Castro~Neto}}\ and\ \bibinfo {author} {\bibfnamefont {F.}~\bibnamefont
  {Guinea}},\ }\href {\doibase 10.1103/PhysRevLett.103.026804} {\bibfield
  {journal} {\bibinfo  {journal} {Phys. Rev. Lett.}\ }\textbf {\bibinfo
  {volume} {103}},\ \bibinfo {pages} {026804} (\bibinfo {year}
  {2009})}\BibitemShut {NoStop}%
\bibitem [{\citenamefont {Young}\ and\ \citenamefont {Kane}(2015)}]{YoungKane}%
  \BibitemOpen
  \bibfield  {author} {\bibinfo {author} {\bibfnamefont {S.~M.}\ \bibnamefont
  {Young}}\ and\ \bibinfo {author} {\bibfnamefont {C.~L.}\ \bibnamefont
  {Kane}},\ }\href {\doibase 10.1103/PhysRevLett.115.126803} {\bibfield
  {journal} {\bibinfo  {journal} {Phys. Rev. Lett.}\ }\textbf {\bibinfo
  {volume} {115}},\ \bibinfo {pages} {126803} (\bibinfo {year}
  {2015})}\BibitemShut {NoStop}%
\bibitem [{\citenamefont {Resta}(2002)}]{Resta2002}%
  \BibitemOpen
  \bibfield  {author} {\bibinfo {author} {\bibfnamefont {R.}~\bibnamefont
  {Resta}},\ }\href {http://stacks.iop.org/0953-8984/14/i=20/a=201} {\bibfield
  {journal} {\bibinfo  {journal} {Journal of Physics: Condensed Matter}\
  }\textbf {\bibinfo {volume} {14}},\ \bibinfo {pages} {R625} (\bibinfo {year}
  {2002})}\BibitemShut {NoStop}%
\bibitem [{\citenamefont {King-Smith}\ and\ \citenamefont
  {Vanderbilt}(1993)}]{Vanderbilt}%
  \BibitemOpen
  \bibfield  {author} {\bibinfo {author} {\bibfnamefont {R.~D.}\ \bibnamefont
  {King-Smith}}\ and\ \bibinfo {author} {\bibfnamefont {D.}~\bibnamefont
  {Vanderbilt}},\ }\href {\doibase 10.1103/PhysRevB.47.1651} {\bibfield
  {journal} {\bibinfo  {journal} {Phys. Rev. B}\ }\textbf {\bibinfo {volume}
  {47}},\ \bibinfo {pages} {1651} (\bibinfo {year} {1993})}\BibitemShut
  {NoStop}%
\bibitem [{\citenamefont {Xie}\ \emph {et~al.}(2014)\citenamefont {Xie},
  \citenamefont {He}, \citenamefont {Chen}, \citenamefont {Feng}, \citenamefont
  {Yi}, \citenamefont {Liang}, \citenamefont {Zhao}, \citenamefont {Mou},
  \citenamefont {He}, \citenamefont {Peng}, \citenamefont {Liu}, \citenamefont
  {Liu}, \citenamefont {Liu}, \citenamefont {Dong}, \citenamefont {Yu},
  \citenamefont {Zhang}, \citenamefont {Zhang}, \citenamefont {Wang},
  \citenamefont {Zhang}, \citenamefont {Yang}, \citenamefont {Peng},
  \citenamefont {Wang}, \citenamefont {Chen}, \citenamefont {Xu},\ and\
  \citenamefont {Zhou}}]{Xie2014}%
  \BibitemOpen
  \bibfield  {author} {\bibinfo {author} {\bibfnamefont {Z.}~\bibnamefont
  {Xie}}, \bibinfo {author} {\bibfnamefont {S.}~\bibnamefont {He}}, \bibinfo
  {author} {\bibfnamefont {C.}~\bibnamefont {Chen}}, \bibinfo {author}
  {\bibfnamefont {Y.}~\bibnamefont {Feng}}, \bibinfo {author} {\bibfnamefont
  {H.}~\bibnamefont {Yi}}, \bibinfo {author} {\bibfnamefont {A.}~\bibnamefont
  {Liang}}, \bibinfo {author} {\bibfnamefont {L.}~\bibnamefont {Zhao}},
  \bibinfo {author} {\bibfnamefont {D.}~\bibnamefont {Mou}}, \bibinfo {author}
  {\bibfnamefont {J.}~\bibnamefont {He}}, \bibinfo {author} {\bibfnamefont
  {Y.}~\bibnamefont {Peng}}, \bibinfo {author} {\bibfnamefont {X.}~\bibnamefont
  {Liu}}, \bibinfo {author} {\bibfnamefont {Y.}~\bibnamefont {Liu}}, \bibinfo
  {author} {\bibfnamefont {G.}~\bibnamefont {Liu}}, \bibinfo {author}
  {\bibfnamefont {X.}~\bibnamefont {Dong}}, \bibinfo {author} {\bibfnamefont
  {L.}~\bibnamefont {Yu}}, \bibinfo {author} {\bibfnamefont {J.}~\bibnamefont
  {Zhang}}, \bibinfo {author} {\bibfnamefont {S.}~\bibnamefont {Zhang}},
  \bibinfo {author} {\bibfnamefont {Z.}~\bibnamefont {Wang}}, \bibinfo {author}
  {\bibfnamefont {F.}~\bibnamefont {Zhang}}, \bibinfo {author} {\bibfnamefont
  {F.}~\bibnamefont {Yang}}, \bibinfo {author} {\bibfnamefont {Q.}~\bibnamefont
  {Peng}}, \bibinfo {author} {\bibfnamefont {X.}~\bibnamefont {Wang}}, \bibinfo
  {author} {\bibfnamefont {C.}~\bibnamefont {Chen}}, \bibinfo {author}
  {\bibfnamefont {Z.}~\bibnamefont {Xu}}, \ and\ \bibinfo {author}
  {\bibfnamefont {X.~J.}\ \bibnamefont {Zhou}},\ }\href
  {https://doi.org/10.1038/ncomms4382} {\bibfield  {journal} {\bibinfo
  {journal} {Nature Communications}\ }\textbf {\bibinfo {volume} {5}},\
  \bibinfo {pages} {3382 EP } (\bibinfo {year} {2014})},\ \bibinfo {note}
  {article}\BibitemShut {NoStop}%
\bibitem [{\citenamefont {Volovik}(2016{\natexlab{b}})}]{Volovik2016}%
  \BibitemOpen
  \bibfield  {author} {\bibinfo {author} {\bibfnamefont {G.~E.}\ \bibnamefont
  {Volovik}},\ }\href {\doibase 10.1134/S0021364016210050} {\bibfield
  {journal} {\bibinfo  {journal} {JETP Letters}\ }\textbf {\bibinfo {volume}
  {104}},\ \bibinfo {pages} {645} (\bibinfo {year}
  {2016}{\natexlab{b}})}\BibitemShut {NoStop}%
\bibitem [{\citenamefont {Jafari}(2019)}]{Jafari2019}%
  \BibitemOpen
  \bibfield  {author} {\bibinfo {author} {\bibfnamefont {S.~A.}\ \bibnamefont
  {Jafari}},\ }\href@noop {} {\bibfield  {journal} {\bibinfo  {journal} {in
  preparation}\ } (\bibinfo {year} {2019})}\BibitemShut {NoStop}%
\bibitem [{\citenamefont {Li}\ \emph {et~al.}(2017)\citenamefont {Li},
  \citenamefont {Rosenstein}, \citenamefont {Shapiro},\ and\ \citenamefont
  {Shapiro}}]{Li}%
  \BibitemOpen
  \bibfield  {author} {\bibinfo {author} {\bibfnamefont {D.}~\bibnamefont
  {Li}}, \bibinfo {author} {\bibfnamefont {B.}~\bibnamefont {Rosenstein}},
  \bibinfo {author} {\bibfnamefont {B.~Y.}\ \bibnamefont {Shapiro}}, \ and\
  \bibinfo {author} {\bibfnamefont {I.}~\bibnamefont {Shapiro}},\ }\href
  {\doibase 10.1103/PhysRevB.95.094513} {\bibfield  {journal} {\bibinfo
  {journal} {Phys. Rev. B}\ }\textbf {\bibinfo {volume} {95}},\ \bibinfo
  {pages} {094513} (\bibinfo {year} {2017})}\BibitemShut {NoStop}%
\bibitem [{\citenamefont {Ezawa}(2012)}]{EzawaPhase}%
  \BibitemOpen
  \bibfield  {author} {\bibinfo {author} {\bibfnamefont {M.}~\bibnamefont
  {Ezawa}},\ }\href {\doibase 10.1103/PhysRevLett.109.055502} {\bibfield
  {journal} {\bibinfo  {journal} {Phys. Rev. Lett.}\ }\textbf {\bibinfo
  {volume} {109}},\ \bibinfo {pages} {055502} (\bibinfo {year}
  {2012})}\BibitemShut {NoStop}%
\bibitem [{\citenamefont {Balakrishnan}\ \emph {et~al.}(2013)\citenamefont
  {Balakrishnan}, \citenamefont {Koon}, \citenamefont {Jaiswal}, \citenamefont
  {Neto},\ and\ \citenamefont {{\"O}zyilmaz}}]{SOI3}%
  \BibitemOpen
  \bibfield  {author} {\bibinfo {author} {\bibfnamefont {J.}~\bibnamefont
  {Balakrishnan}}, \bibinfo {author} {\bibfnamefont {G.~K.~W.}\ \bibnamefont
  {Koon}}, \bibinfo {author} {\bibfnamefont {M.}~\bibnamefont {Jaiswal}},
  \bibinfo {author} {\bibfnamefont {A.~C.}\ \bibnamefont {Neto}}, \ and\
  \bibinfo {author} {\bibfnamefont {B.}~\bibnamefont {{\"O}zyilmaz}},\ }\href
  {https://www.nature.com/articles/nphys2576} {\bibfield  {journal} {\bibinfo
  {journal} {Nature Physics}\ }\textbf {\bibinfo {volume} {9}},\ \bibinfo
  {pages} {284} (\bibinfo {year} {2013})}\BibitemShut {NoStop}%
\bibitem [{\citenamefont {Bradley}\ and\ \citenamefont
  {Cracknell}(2010)}]{Bradley}%
  \BibitemOpen
  \bibfield  {author} {\bibinfo {author} {\bibfnamefont {C.}~\bibnamefont
  {Bradley}}\ and\ \bibinfo {author} {\bibfnamefont {A.}~\bibnamefont
  {Cracknell}},\ }\href@noop {} {\emph {\bibinfo {title} {The mathematical
  theory of symmetry in solids: representation theory for point groups and
  space groups}}}\ (\bibinfo  {publisher} {Oxford University Press},\ \bibinfo
  {year} {2010})\BibitemShut {NoStop}%
\bibitem [{\citenamefont {Bunkar}(1979)}]{BunkarChemist}%
  \BibitemOpen
  \bibfield  {author} {\bibinfo {author} {\bibfnamefont {P.~R.}\ \bibnamefont
  {Bunkar}},\ }\href@noop {} {\emph {\bibinfo {title} {Molecular Symmetry and
  Spectroscopy}}}\ (\bibinfo  {publisher} {ACADEMIC PRESS},\ \bibinfo {year}
  {1979})\BibitemShut {NoStop}%
\bibitem [{\citenamefont {Kittel}(1983)}]{Kittel}%
  \BibitemOpen
  \bibfield  {author} {\bibinfo {author} {\bibfnamefont {C.}~\bibnamefont
  {Kittel}},\ }\href@noop {} {\emph {\bibinfo {title} {Quantum theory of
  solids, 2nd Revised Edition}}}\ (\bibinfo  {publisher} {John Wiley and
  Sons},\ \bibinfo {year} {1983})\BibitemShut {NoStop}%
\bibitem [{\citenamefont {Elcoro}\ \emph {et~al.}(2017)\citenamefont {Elcoro},
  \citenamefont {Bradlyn}, \citenamefont {Wang}, \citenamefont {Vergniory},
  \citenamefont {Cano}, \citenamefont {Felser}, \citenamefont {Bernevig},
  \citenamefont {Orobengoa}, \citenamefont {Flor},\ and\ \citenamefont
  {Aroyo}}]{Bilbao}%
  \BibitemOpen
  \bibfield  {author} {\bibinfo {author} {\bibfnamefont {L.}~\bibnamefont
  {Elcoro}}, \bibinfo {author} {\bibfnamefont {B.}~\bibnamefont {Bradlyn}},
  \bibinfo {author} {\bibfnamefont {Z.}~\bibnamefont {Wang}}, \bibinfo {author}
  {\bibfnamefont {M.~G.}\ \bibnamefont {Vergniory}}, \bibinfo {author}
  {\bibfnamefont {J.}~\bibnamefont {Cano}}, \bibinfo {author} {\bibfnamefont
  {C.}~\bibnamefont {Felser}}, \bibinfo {author} {\bibfnamefont {B.~A.}\
  \bibnamefont {Bernevig}}, \bibinfo {author} {\bibfnamefont {D.}~\bibnamefont
  {Orobengoa}}, \bibinfo {author} {\bibfnamefont {G.}~\bibnamefont {Flor}}, \
  and\ \bibinfo {author} {\bibfnamefont {M.~I.}\ \bibnamefont {Aroyo}},\
  }\href@noop {} {\bibfield  {journal} {\bibinfo  {journal} {Journal of Applied
  Crystallography}\ }\textbf {\bibinfo {volume} {50}},\ \bibinfo {pages} {1457}
  (\bibinfo {year} {2017})}\BibitemShut {NoStop}%
\bibitem [{\citenamefont {Bradlyn}\ \emph {et~al.}(2017)\citenamefont
  {Bradlyn}, \citenamefont {Elcoro}, \citenamefont {Cano}, \citenamefont
  {Vergniory}, \citenamefont {Wang}, \citenamefont {Felser}, \citenamefont
  {Aroyo},\ and\ \citenamefont {Bernevig}}]{EBRnature}%
  \BibitemOpen
  \bibfield  {author} {\bibinfo {author} {\bibfnamefont {B.}~\bibnamefont
  {Bradlyn}}, \bibinfo {author} {\bibfnamefont {L.}~\bibnamefont {Elcoro}},
  \bibinfo {author} {\bibfnamefont {J.}~\bibnamefont {Cano}}, \bibinfo {author}
  {\bibfnamefont {M.}~\bibnamefont {Vergniory}}, \bibinfo {author}
  {\bibfnamefont {Z.}~\bibnamefont {Wang}}, \bibinfo {author} {\bibfnamefont
  {C.}~\bibnamefont {Felser}}, \bibinfo {author} {\bibfnamefont
  {M.}~\bibnamefont {Aroyo}}, \ and\ \bibinfo {author} {\bibfnamefont {B.~A.}\
  \bibnamefont {Bernevig}},\ }\href@noop {} {\bibfield  {journal} {\bibinfo
  {journal} {Nature}\ }\textbf {\bibinfo {volume} {547}},\ \bibinfo {pages}
  {298} (\bibinfo {year} {2017})}\BibitemShut {NoStop}%
\bibitem [{\citenamefont {Cano}\ \emph {et~al.}(2018)\citenamefont {Cano},
  \citenamefont {Bradlyn}, \citenamefont {Wang}, \citenamefont {Elcoro},
  \citenamefont {Vergniory}, \citenamefont {Felser}, \citenamefont {Aroyo},\
  and\ \citenamefont {Bernevig}}]{EBRCano}%
  \BibitemOpen
  \bibfield  {author} {\bibinfo {author} {\bibfnamefont {J.}~\bibnamefont
  {Cano}}, \bibinfo {author} {\bibfnamefont {B.}~\bibnamefont {Bradlyn}},
  \bibinfo {author} {\bibfnamefont {Z.}~\bibnamefont {Wang}}, \bibinfo {author}
  {\bibfnamefont {L.}~\bibnamefont {Elcoro}}, \bibinfo {author} {\bibfnamefont
  {M.}~\bibnamefont {Vergniory}}, \bibinfo {author} {\bibfnamefont
  {C.}~\bibnamefont {Felser}}, \bibinfo {author} {\bibfnamefont
  {M.}~\bibnamefont {Aroyo}}, \ and\ \bibinfo {author} {\bibfnamefont {B.~A.}\
  \bibnamefont {Bernevig}},\ }\href@noop {} {\bibfield  {journal} {\bibinfo
  {journal} {Physical Review B}\ }\textbf {\bibinfo {volume} {97}},\ \bibinfo
  {pages} {035139} (\bibinfo {year} {2018})}\BibitemShut {NoStop}%
\bibitem [{\citenamefont {Bradlyn}\ \emph {et~al.}(2018)\citenamefont
  {Bradlyn}, \citenamefont {Elcoro}, \citenamefont {Vergniory}, \citenamefont
  {Cano}, \citenamefont {Wang}, \citenamefont {Felser}, \citenamefont {Aroyo},\
  and\ \citenamefont {Bernevig}}]{EBRbradlyn}%
  \BibitemOpen
  \bibfield  {author} {\bibinfo {author} {\bibfnamefont {B.}~\bibnamefont
  {Bradlyn}}, \bibinfo {author} {\bibfnamefont {L.}~\bibnamefont {Elcoro}},
  \bibinfo {author} {\bibfnamefont {M.}~\bibnamefont {Vergniory}}, \bibinfo
  {author} {\bibfnamefont {J.}~\bibnamefont {Cano}}, \bibinfo {author}
  {\bibfnamefont {Z.}~\bibnamefont {Wang}}, \bibinfo {author} {\bibfnamefont
  {C.}~\bibnamefont {Felser}}, \bibinfo {author} {\bibfnamefont
  {M.}~\bibnamefont {Aroyo}}, \ and\ \bibinfo {author} {\bibfnamefont {B.~A.}\
  \bibnamefont {Bernevig}},\ }\href@noop {} {\bibfield  {journal} {\bibinfo
  {journal} {Physical Review B}\ }\textbf {\bibinfo {volume} {97}},\ \bibinfo
  {pages} {035138} (\bibinfo {year} {2018})}\BibitemShut {NoStop}%
\bibitem [{\citenamefont {Kajita}\ \emph
  {et~al.}(2014{\natexlab{b}})\citenamefont {Kajita}, \citenamefont {Nishio},
  \citenamefont {Tajima}, \citenamefont {Suzumura},\ and\ \citenamefont
  {Kobayashi}}]{Organic3}%
  \BibitemOpen
  \bibfield  {author} {\bibinfo {author} {\bibfnamefont {K.}~\bibnamefont
  {Kajita}}, \bibinfo {author} {\bibfnamefont {Y.}~\bibnamefont {Nishio}},
  \bibinfo {author} {\bibfnamefont {N.}~\bibnamefont {Tajima}}, \bibinfo
  {author} {\bibfnamefont {Y.}~\bibnamefont {Suzumura}}, \ and\ \bibinfo
  {author} {\bibfnamefont {A.}~\bibnamefont {Kobayashi}},\ }\href {\doibase
  10.7566/JPSJ.83.072002} {\bibfield  {journal} {\bibinfo  {journal} {Journal
  of the Physical Society of Japan}\ }\textbf {\bibinfo {volume} {83}},\
  \bibinfo {pages} {072002} (\bibinfo {year} {2014}{\natexlab{b}})}\BibitemShut
  {NoStop}%
\bibitem [{\citenamefont {Liu}\ \emph {et~al.}(2010)\citenamefont {Liu},
  \citenamefont {Qi}, \citenamefont {Zhang}, \citenamefont {Dai}, \citenamefont
  {Fang},\ and\ \citenamefont {Zhang}}]{Liu2010}%
  \BibitemOpen
  \bibfield  {author} {\bibinfo {author} {\bibfnamefont {C.-X.}\ \bibnamefont
  {Liu}}, \bibinfo {author} {\bibfnamefont {X.-L.}\ \bibnamefont {Qi}},
  \bibinfo {author} {\bibfnamefont {H.}~\bibnamefont {Zhang}}, \bibinfo
  {author} {\bibfnamefont {X.}~\bibnamefont {Dai}}, \bibinfo {author}
  {\bibfnamefont {Z.}~\bibnamefont {Fang}}, \ and\ \bibinfo {author}
  {\bibfnamefont {S.-C.}\ \bibnamefont {Zhang}},\ }\href {\doibase
  10.1103/PhysRevB.82.045122} {\bibfield  {journal} {\bibinfo  {journal} {Phys.
  Rev. B}\ }\textbf {\bibinfo {volume} {82}},\ \bibinfo {pages} {045122}
  (\bibinfo {year} {2010})}\BibitemShut {NoStop}%
\end{thebibliography}%
\end{document}